\documentclass[%
reprint,
superscriptaddress,
showpacs,
nofootinbib,
amsmath,amssymb,
prc,
floatfix, ]%
{revtex4-1}

\usepackage{color}

\usepackage{graphicx}
\usepackage{dcolumn}
\usepackage{bm}
\usepackage[bookmarks=true,colorlinks,%
            citecolor=blue,linkcolor=blue,anchorcolor=blue,filecolor=blue,urlcolor=blue,%
           ]{hyperref}

\allowdisplaybreaks

\newcommand{\bra}[1]{\langle {#1} |}
\newcommand{\ket}[1]{| {#1} \rangle}
\newcommand{\inproduct}[2]{\langle #1 | #2 \rangle}

\begin{document}


\title{
Microscopic collective inertial masses for nuclear reaction
	in the presence of nucleonic effective mass}

\author{Kai Wen}%
 \email{wenkai@nucl.ph.tsukuba.ac.jp}
 \affiliation{Center for Computational Sciences,
              University of Tsukuba, Tsukuba 305-8577, Japan}

\author{Takashi Nakatsukasa}%
 \affiliation{Center for Computational Sciences,
              University of Tsukuba, Tsukuba 305-8577, Japan}
 \affiliation{Faculty of Pure and Applied Sciences,
              University of Tsukuba, Tsukuba 305-8571, Japan}
 \affiliation{RIKEN Nishina Center, Wako 351-0198, Japan}

\date{\today}

\begin{abstract}

Collective inertial mass coefficients with respect to translational,
relative, and rotational motions are microscopically calculated,
along the collective reaction path self-consistently determined,
based on
the adiabatic self-consistent collective coordinate (ASCC) method.
The impact of the time-odd component of the mean-field potential on the
inertial masses are investigated.
The results are compared with those calculated with the cranking formulae.
The inertial masses based on the ASCC method reproduce the exact
total nuclear mass for the translational motion
as well as the exact reduced
masses as the asymptotic values
for the relative and rotational motions.
In contrast, the cranking formulae fail to do so.
This is due to the fact that the (local) Galilean invariance
is properly restored in the ASCC
method, but violated in the cranking formulae.
A model Hamiltonian for low-energy nuclear reaction is
constructed with
the microscopically derived potentials and inertial masses.
The astrophysical $S$-factors are calculated,
which indicates
the importance
of microscopic calculation of
proper inertial masses.

\end{abstract}


\maketitle


\section{Introduction}

Theoretical description of low-energy nuclear reaction with a
solid microscopic foundation is still a challenging subject in
nuclear physics.
It may provide us with a deep insight into the reaction mechanism
and the quantum dynamics of many-nucleon systems.

Starting from a microscopic many-body Hamiltonian (an energy density
functional),
the time-dependent Hartree-Fock (TDHF) method \cite{Neg82,Sim12,MRSU14}
has been established as a useful tool to
study low-energy nuclear reactions.
The TDHF real-time simulations have been intensively carried out with
significant developments
\cite{NY05,Sim12,NMMY16,SY16,UOS16,Sek17,USY17},
including the time-dependent Hartree-Fock-Bogoliubov (TDHFB) calculations
\cite{Eba10,SBMR11,Has12,Has13,HS16,NMMY16,BMRS16,MSW17,SH19}.
In the TDHF(B) simulations, the reaction path is given by
a time-dependent generalized Slater determinant
which is determined by the TDHF(B) equations as an initial-value problem.
It is successful
in describing the microscopic dynamics in major reaction channels on average.
However, it is known that
important quantum effects are missing in the TDHF(B).
For instance the sub-barrier fusion reaction and
spontaneous fission
can not be properly described
within the real-time TDHF(B) simulations \citep{RS80,Neg82,NMMY16}.

To recover the quantum fluctuations of
nuclear collective dynamics that is missing in TDHF(B),
we aim at the requantization of TDHF(B) on the collective subspace \cite{NMMY16}.
First, we find a collective subspace spanned by a few selected
collective canonical variables, which is well decoupled from the intrinsic excitations.
The collective subspace and the collective variables can
be extracted using the adiabatic self-consistent
collective coordinate (ASCC) method \cite{MNM00,HNMM07,HNMM09,Nak12,NMMY16}.
The ASCC is derived
from the invariance principle of the TDHF(B) equation \cite{MMSK80} under an assumption
of adiabatic (slow) collective motion,
namely, second-order expansion with respect to the collective momenta \cite{MNM00}.
It also has a close connection with a theory of large amplitude collective motion
developed in Refs.~\cite{KWD91,DKW00}.
The collective subspace in the adiabatic regime is given by
a series of time-even Slater determinants and generators of
collective coordinates and momenta locally defined at each state.
In the present paper, we apply the method to the low-energy nuclear reaction,
to identify the optimal reaction path and the canonical variables.
The requantization is performed on
this self-consistently determined collective reaction path.

One of the key ingredients of the requantization procedure
is the inertial mass parameters with respect
to the collective coordinates.
This is essential for the construction of the collective Hamiltonian.
For the nuclear reaction,
the relative distance $R$ between two colliding nuclei
should be a proper choice of the collective coordinate
in the asymptotic region ($R\rightarrow\infty$).
The corresponding inertial mass should be the reduced mass $\mu_{\rm red}=A_P A_T m/(A_P+A_T)$,
where $A_P$ ($A_T$) is the mass number
of the projectile (target) nucleus
and $m$ is the bare nucleon mass.
Therefore, a
theory can be tested by examining its asymptotic limit.
On the other hand, the inertial mass in the interior region where two nuclei touch
each other is highly nontrivial.
Thus, a microscopic theory to calculate the mass on the entire reaction path is necessary.

A commonly used method to calculate the collective mass
coefficient is the cranking formula \citep{BK68-1,BK68-2,YLQG99} which is
derived based on the adiabatic perturbation theory.
However, it is known that the cranking formula
fails to reproduce the total mass
of a whole nucleus for the translational motion,
when the mean-field potential has velocity dependence \cite{RS80}.
This is due to violation of the Galilean invariance
in the mean-field level
which must be corrected by residual fields caused by the translational motion \cite{BM75}.
From this failure, we may expect that the cranking formula cannot reproduce the reduced mass
for the relative motion of two nuclei at $R\rightarrow \infty$.
Since most of the realistic nuclear energy density functionals give effective mass,
$m^*/m<1$, the mean-field potential indeed has the velocity dependence.
Therefore, it is highly desirable to find a proper method for the
calculation of inertial masses.
In this paper, we show that
the ASCC method, which is capable of taking into account the residual effects
missing in the cranking formula, provides a promising tool
to microscopically calculate the proper inertial masses.

In our previous works \cite{WenN16,WenN17}, we calculated
the ASCC inertial masses for the relative motion between two nuclei,
for the velocity-independent mean-field potential.
We also examined those of the cranking formula, which turned out to be
almost identical to the ASCC mass at $R\rightarrow\infty$.
However, this is not true in general.
In this paper, we show that the asymptotic value in the cranking formula
for a velocity-dependent potential does not agree with the reduced mass,
while that in the ASCC method exactly reproduces it.

In general, the non-local mean-field potential produces the effective mass $m^*/m\neq 1$.
For the energy density functionals of the Skyrme type,
the effective mass comes from the $t_1$ and $t_2$ terms
(momentum-dependent terms of the Skyrme interaction).
In order to guarantee the Galilean invariance of the energy density functional,
we need densities that are time-odd with respect to
the time-reversal transformation.
These time-odd densities vanish in the ground state of even-even nuclei.
Therefore they do not contribute to the static mean-field potential.
However, the time-odd mean fields may appear and play important roles
in time-dependent dynamics, because the time-odd densities are non-zero in general
for the time-dependent states.
In this paper, we investigate the impact of the time-odd mean fields
on the collective inertial masses
in the context of low-energy reaction dynamics of light nuclei.

The paper is organized as follows: In Sec.~\ref{sec:theo}, we recapitulate
the theoretical methods, mainly the ASCC method and cranking formulae,
for the calculation of inertial masses. In Sec.~\ref{sec:application},
in the presence of time-odd mean-field potential, we apply different methods
to calculate the mass coefficients for the
translational, relative and rotational motions,
the applications are carried out for the reaction systems of
$\alpha+\alpha\rightarrow$ $^{8}$Be,
$\alpha+^{16}$O $\rightarrow$ $^{20}$Ne
and $^{16}$O$+^{16}$O $\rightarrow$ $^{32}$S.
The impacts of inertial masses on the
$S$-factors of sub-barrier fusion
are discussed.
A summary and concluding remarks are given in Sec.~\ref{sec:sum}.

\section{\label{sec:theo}Theoretical methods}


It is customary to use a model Hamiltonian for studies of low-energy nuclear reactions.
The relative distance $R$ between the projectile and the target nuclei
is assumed to be a dynamical coordinate, then, the Schr\"odinger equation is given in
a form,
\begin{equation}
	\left\{	-\frac{1}{2\mu_{\rm red}}\frac{d^2}{dR^2}
	+ \frac{\ell(\ell+1)}{2\mu_{\rm red}R^2}+V(R) -E \right\}
	u_\ell (R) = 0 ,
	\label{reaction_model}
\end{equation}
where $\mu_{\rm red}$ is the reduced mass,
$\mu_{\rm red}=A_P A_T m/(A_P+A_T)$.
We use the unit $\hbar=1$ throughout this paper.
Here, in addition to the assumption of the coordinate $R$,
the two nuclei are approximated as structureless ``point'' particles,
which leads to the trivial inertial masses,
$\mu_{\rm red}$ and $\mu_{\rm red}R^2$.
This approximation becomes exact in the asymptotic limit ($R\rightarrow\infty$).
However, at finite distance $R$, even if the two nuclei are well apart,
it is not trivial to justify the point-particle approximation.
In fact, as we show later, the rigid-body values of the moments of inertia
are significantly different from those of the point-particle approximation,
${\cal J}_{\rm rigid}\neq \mu_{\rm red}R^2$, even at $R>R_P+R_T$
where $R_P$ ($R_T$) indicates the radius of the projectile (target) nucleus.
It should be noted that the moments of inertia for deformed nuclei are given
by the rigid-body values in the harmonic-oscillator-potential model \cite{BM75}.
Therefore, it is of great interest to examine how far the point-particle approximation
can be justified.

To reveal the coordinate ($R$) dependence and dynamical properties
of the inertial masses,
we apply the ASCC method to an energy density functional.
In this section we recapitulate the ASCC method for the calculations of
inertial masses.
For comparison, we also give a brief description of the cranking formulae.

\subsection{\label{sec:theoa1}ASCC inertial masses}

The ASCC method provides a collective subspace which describes nuclear reaction.
The generators of the collective variables $(q,p)$ are locally given by
$(\hat{Q}(q),\hat{P}(q))$ and determined by the ASCC equations.
In the asymptotic region on the reaction path (collective subspace) at $p=0$,
these should correspond to the relative distance and its conjugate momentum.
Thus, we expect that
\begin{equation}
	\hat{Q}(q)\ket{\Psi(q)}\propto \hat{R} \ket{\Psi(q)},
	\quad
	\hat{P}(q)\ket{\Psi(q)}\propto \frac{\partial}{\partial R} \ket{\Psi(q)} ,
	\label{asymptotic_limit}
\end{equation}
where the proportional constants are arbitrary as far as they
satisfy the weak canonicity condition
\begin{eqnarray}
 \langle\Psi(q)|[i\hat{P}(q),\hat{Q}(q)]|\Psi(q)\rangle=1 . \label{weak}
\end{eqnarray}
In this paper, we study systems that preserve the axial symmetry on the reaction path.
The one-body operator $\hat{R}$ for the relative distance
between projectile and target nuclei can be defined in the following way.
The symmetry axis is chosen as the $z$ axis,
assuming the projectile on the right ($z>0$) and target on the left ($z<0$).
We introduce a separation plane at $z=z_s$, so that
\begin{eqnarray}
	\int d\mathbf{r}\ \theta(z-z_s) \rho(\mathbf{r}) = A_P ,
\end{eqnarray}
where $\rho(\mathbf{r})$ is the nucleon density and
$\theta(x)$ is the step function.
The operator $\hat{R}$ is given by $\hat{R}\equiv \hat{Z}_P-\hat{Z}_T$, with
\begin{equation}
	\hat{Z}_{P(T)}=\sum_{s=\uparrow,\downarrow}\sum_{q=n,p} \int d\mathbf{r}
	\hat{\psi}_{sq}^\dagger(\mathbf{r})\hat{\psi}_{sq}(\mathbf{r})
	 z \frac{\theta(\pm(z-z_{\rm s}))}{A_{P(T)}} ,
\end{equation}
where the upper sign is adopted for $\hat{Z}_P$ and the lower for $\hat{Z}_T$.

The operator $\hat{R}$ is artificially defined as a convenient measure
for the relative distance.
Its meaning is clear and well defined in the asymptotic region,
though it is no longer meaningful
when the projectile and the target are merged into a single nucleus.
Nevertheless, this choice of $\hat{R}$ does not affect the final result
as far as there is a one-to-one correspondence between $q$ and $R$,
because the reaction path $\ket{\Psi(q)}$ and its canonical variables $(q,p)$ are
self-consistently determined without any assumption.
After determining all the quantities in the collective Hamiltonian,
we perform the coordinate transformation from $q$ to $R$,
by using the relation $R(q)=\bra{\Psi(q)}\hat{R}\ket{\Psi(q)}$.
This is merely a change in the scale (and the dimension) of the coordinate.

In the ASCC method, the reaction path $\ket{\Psi(q)}$ at $p=0$
and the local generators
$(\hat{Q}(q),\hat{P}(q))$ are determined by solving
the ASCC equations
\begin{eqnarray}
&&\delta\langle \Psi(q)
|\hat{H}_{\rm mv}|
\Psi(q)\rangle = 0, \label{chf}\\
&&\delta\langle \Psi(q)|[\hat{H}_{\rm mv},\frac{1}{i}\hat{P}(q) ]
        - \frac{\partial^{2} V(q)}{\partial q^{2}} \hat{Q}(q)
        |\Psi(q)\rangle = 0, \label{RPA0} \\
&&\delta\langle \Psi(q)|[\hat{H}_{\rm mv},i\hat{Q}(q)]
         - \frac{1}{M(q)}\hat{P}(q)   |\Psi(q)\rangle = 0, \label{RPA}
\end{eqnarray}
where curvature terms are neglected for simplicity \cite{MNM00}.
The Hamiltonian in the moving frame $\hat{H}_{\rm mv}$ and the potential
$V(q)$ are respectively defined as
\begin{eqnarray}
\hat{H}_{\rm mv}\equiv \hat{H}- \frac{\partial V(q)}{\partial q} \hat{Q}(q),\quad
V(q)\equiv \langle \Psi(q)|\hat{H} |\Psi(q)\rangle. \label{pdef}
\end{eqnarray}
$V(q)$ in Eq. (\ref{pdef}) is used as $V(R)$ in Eq. (\ref{reaction_model})
at $R=\bra{\Psi(q)}\hat{R}\ket{\Psi(q)}$.

Equation (\ref{chf}) which depends on $\hat{Q}(q)$
determines the reaction path $\ket{\Psi(q)}$.
The local generators, $\hat{P}(q)$ and $\hat{Q}(q)$, are
given by a solution of Eqs. (\ref{RPA0}), and (\ref{RPA}).
However, since solutions of Eqs. (\ref{RPA0}) and (\ref{RPA}) are not unique,
we need to select the one corresponding to the reaction path we study.
Normally, we choose a solution with the lowest frequency,
$\omega^2 = (\partial^2 V/\partial q^2)M^{-1}(q)$, except for the Nambu-Goldstone (NG) modes.
The generators should satisfy the asymptotic condition of Eq. (\ref{asymptotic_limit}).

In Eq. (\ref{RPA}) there appears the inertial mass parameter $M(q)$
with respect to the coordinate $q$.
The magnitude of $M(q)$ depends on the scale of the
coordinate $q$ which is arbitrary.
We can choose to set the mass $M(q)$ in Eq. (\ref{RPA})
to be a constant, for instance
$M_q=1$ MeV$^{-1}$$[q]^{-2}$
without losing generality~\cite{WenN16},
where $[q]$ represents the unit of $q$.

In order to obtain an intuitive picture of the collective
dynamics,
we map the collective coordinate $q$ to
the relative distance $R$ between projectile and target nuclei.
The inertial mass should be transformed as
\begin{eqnarray}
	M(R) =M(q)\left(\frac{dq}{dR}\right)^{2}
	= M(q) \left(\frac{dR}{dq}\right)^{-2},
\label{mass}
\end{eqnarray}
where the derivative $dR/dq$ can be obtained as
\begin{eqnarray}
\frac{dR}{dq} &=&
\frac{d}{dq}\langle \Psi(q) |\hat{R}| \Psi(q) \rangle
 =\langle \Psi(q) | [\hat{R},\frac{1}{i}\hat{P}(q) ] | \Psi(q) \rangle
 \nonumber \\
 &=&2\sum_{n\in p, j\in h} R_{nj}(q)P_{nj}(q)
,\label{mass3}
\end{eqnarray}
with the local generator $\hat{P}(q)$.
$R_{nj}(q)$ and
$P_{nj}(q)$
are the particle-hole (ph) matrix
elements of $\hat{R}(q)$ and $\hat{P}(q)$,
obtained by solving ASCC equations (\ref{RPA0}) and (\ref{RPA})
\cite{WenN16}.
The calculated mass $M(R)$ will replace the constant mass $\mu_{\rm red}$
in the first term of Eq. (\ref{reaction_model}).

Apart from the the relative motion of our current interest,
Eqs. (\ref{RPA0}) and (\ref{RPA}) provide solutions for the NG modes,
such as the translation and the rotation.
These modes have natural and global generators, namely,
the total linear momentum $\hat{\mathbf{P}}$ for the translation
and the total angular momentum $\hat{\mathbf{J}}$ for the rotation.
For example, the rotational motion around the $x$ axis can be generated
by $\hat{J}_x$ without any cost of energy ($\omega^2=0$),
leading to the Thouless-Valatin equations \cite{TV62,RS80},
\begin{eqnarray}
	&&\delta\bra{\Psi(q)} \left[\hat{H}_{\rm mv},\hat{J}_x \right] \ket{\Psi(q)}= 0 ,
	\label{RPA0-2}\\
	&&\delta\bra{\Psi(q)} \left[\hat{H}_{\rm mv},i\hat{\Theta}(q)\right]
	 - \frac{1}{{\cal J}(q)}\hat{J}_x   |\Psi(q)\rangle = 0 ,
	 \label{RPA-2}\\
	&&\bra{\Psi(q)}[i\hat{J}_x, \hat{\Theta}(q)] \ket{\Psi(q)}=1,
\end{eqnarray}
where $\hat{\Theta}$ is the angle variable conjugate to $\hat{J}_x$, and
${\cal J}(q)$ is the moment of inertia around $x$ axis.
Equations (\ref{RPA0-2}) and (\ref{RPA-2}) indicate that
$(\hat{Q}(q),\hat{P}(q))=(\hat{\Theta}(q),\hat{J}_x)$
correspond to a solution of the ASCC equations (\ref{RPA0}) and (\ref{RPA})
with $\partial^2 V/\partial q^2=0$ and $M(q)={\cal J}(q)$.
The calculated rotational moments of inertia can be used
to replace $\mu_{\rm red}R^2$
in the second term of Eq. (\ref{reaction_model}).

To calculate the inertial masses for the NG modes,
an efficient method was proposed in Ref. \cite{Hin15}.
The inertial mass of the NG modes $M(q)$ are given by
the zero-frequency limit of the strength function in the linear response
for the momentum operator $\hat{P}$.
We apply this method to
	the calculations
of the translational
and rotational inertial masses.

\subsection{\label{sec:CHF+cranking}CHF+cranking method}

A simple method based on the cranking formula to calculate the inertial mass
is widely used for nuclear collective motion.
The collective path (reaction path) is usually produced by
the constrained-Hartree-Fock (CHF)
calculation with a constraining operator $\hat{C}$ given by hand.
The CHF states are given by the variation
\begin{equation}
	\delta \bra{\Psi_0(\lambda)}\hat{H}_\lambda\ket{\Psi_0(\lambda)}=0,
	\quad
	\hat{H}_\lambda \equiv\hat{H}-\lambda \hat{C} ,
\end{equation}
where $\lambda$ is the Lagrange multiplier.
Upon the CHF states $\ket{\Psi_0(R)}=\ket{\Psi_0(\lambda)}$ with
$R=\bra{\Psi_0(\lambda)}\hat{R}\ket{\Psi_0(\lambda)}$,
the inertial mass with respect to the coordinate $R$
is calculated using the cranking formula \cite{Bar11}:
\begin{eqnarray}
M_{\rm cr}^{\rm np}(R)
	&=&2 \sum_{m}
	\frac{|\bra{\Psi_m(R)} \partial/\partial R \ket{\Psi_0(R)}|^2}{E_m(R)-E_0(R)} \\
	&=&2 \sum_{n\in p,j\in h}
\frac{|\bra{\varphi_n(R)}
	\partial/\partial R
	\ket{\varphi_j(R)}|^2}
{e_n(R)-e_j(R)} ,
\label{NP_cranking}
\end{eqnarray}
where the single-particle states $\ket{\varphi_\mu(R)}$ and energies $e_\mu(R)$
are defined with respect to the CHF single-particle Hamiltonian,
$\hat{h}_{\lambda}=\hat{h}_{\rm HF}[\rho]-\lambda \hat{C}$, as
\begin{equation}
	\hat{h}_\lambda\ket{\varphi_\mu(R)}=e_\mu(R))\ket{\varphi_\mu(R)} ,
	\label{CHF_eq}
\quad \mu \in p, h .
\end{equation}
Note that the Lagrange multiplier $\lambda$ is a function of $R$ given by
the condition, $R=\bra{\Psi_0(\lambda)}\hat{R}\ket{\Psi_0(\lambda)}$.
The constraining operator $\hat{C}$ can be in general different from
the relative-distance
operator
$\hat{R}$.
In the following calculations in Sec.~\ref{sec:application},
we use the mass quadrupole and octupole
operators
as $\hat{C}$.



Another cranking formula, which is even more frequently used in many applications,
can be derived by neglecting the rearranged fields
induced by the change of $\lambda$,
namely, $\delta h_{\rm HF}/\delta\rho \cdot d\rho/d\lambda=0$.
Normally, one chooses the constraining operator as the collective coordinate,
$\hat{C}=\hat{R}$, for the CHF calculation.
However, in the present calculation,
since we need to find the separation plane $z=z_s$ to determine
the operator $\hat{R}$,
it is convenient to adopt a different one-body operator $\hat{C}$.
Let us derive the formula for this case.
From Eq. (\ref{CHF_eq}) and the orthonormality condition
$\inproduct{\varphi_\mu(R)}{\varphi_\nu(R)}=\delta_{\mu\nu}$,
it is easy to find
\begin{equation}
	\bra{\varphi_n}\frac{\partial}{\partial R}\ket{\varphi_j}
	= - \frac{\bra{\varphi_n}\partial h_\lambda /\partial R \ket{\varphi_j}}
	{e_n-e_j}
	\approx \frac{d\lambda}{dR}\frac{\bra{\varphi_n}\hat{C}\ket{\varphi_j}}
	{e_n-e_j} .
	\label{cranking_approximation}
\end{equation}
Neglecting the rearrangement,
$\hat{h}_{\lambda+\Delta\lambda}\approx \hat{h}_\lambda-\Delta\lambda\hat{C}$,
the derivative $dR/d\lambda$ is estimated with the first-order perturbation as
${dR}/{d\lambda}= 2 S_1 (\hat{R},\hat{C})$,
where
\begin{eqnarray}
	S_k (\hat{Q}_1,\hat{Q}_2)&\equiv&
        \frac{1}{2}
	\sum_{m>0}
	\frac{\bra{\Psi_0}\hat{Q}_1\ket{\Psi_m}
	\bra{\Psi_m}\hat{Q}_2\ket{\Psi_0}+\mbox{c.c.}}
	{\left(E_{m}-E_0\right)^k} \nonumber\\
        &=& \frac{1}{2}
\sum_{n\in p,j\in h}\frac{\bra{\varphi_j}\hat{Q}_1\ket{\varphi_n}
\bra{\varphi_n}\hat{Q}_2\ket{\varphi_j}+\mbox{c.c.}}
	{\left(e_{n}-e_{j}\right)^k}  \nonumber \\
&=& S_k (\hat{Q}_2,\hat{Q}_1) .
\label{S_k}
\end{eqnarray}
Thus, the formula reads
\begin{equation}
M_{\rm cr}^{\rm p}(R)=
\frac{1}{2}
	S_3(\hat{C},\hat{C}) \left\{ S_1(\hat{R},\hat{C})\right\}^{-2} .
\label{P_cranking}
\end{equation}
According to Ref.~\cite{Bar11},
we call the former one in Eq. (\ref{NP_cranking})
``non-perturbative'' cranking inertia and the latter in Eq. (\ref{P_cranking})
``perturbative'' one.
In contrast to the ASCC mass,
the cranking masses of Eqs. (\ref{NP_cranking}) and (\ref{P_cranking}) have
a drawback that they
both neglect residual effect of the time-odd density fluctuation.
As we shall see in Sec.~\ref{sec:application},
when the nucleonic effective masses are present,
the cranking formulae produce wrong masses for the translation
and for the relative motion at $R\rightarrow\infty$.

For the NG modes such as the translation and the rotation,
we know the generator of the collective coordinate.
For instance, in the translational case,
the center-of-mass (COM) coordinates $X_k$ ($k=x,y,z$) are the trivial collective coordinates.
Then, the non-perturbative cranking formula (\ref{NP_cranking})
can be transformed into
\begin{eqnarray}
M_{\rm cr}^{\rm np}(X_k)&=&2 \sum_{n\in p,j\in h}
\frac{|\bra{\varphi_n(X_k)}\partial/\partial X_k \ket{\varphi_j(X_k)}|^2}
{e_n(X_k)-e_j(X_k)} \\
&=&2 \sum_{n\in p,j\in h}
	\frac{|\bra{\varphi_n} \hat{p}_k \ket{\varphi_j}|^2}{e_n-e_j} ,
	\label{cranking_translation}
\end{eqnarray}
where $\hat{p}_k$ is the single-particle linear momentum.
In the second equation,
we take advantage of
the fact that the single-particle energies and the wave functions
relative to the COM are invariant with respect to its position $X_k$.
For the rotation, we may replace $X_k$ by an angle $\phi_k$
and $\hat{p}_k$ by the angular momentum $\hat{j}_k$.
This leads to nothing but the cranking formula for
the moment of inertia, originally proposed by Inglis \cite{Ing54},
\begin{equation}
\left({\cal J}_{\rm cr}^{\rm np}\right)_k =2 \sum_{n\in p,j\in h}
	\frac{|\bra{\varphi_n} \hat{j}_k \ket{\varphi_j}|^2}{e_n-e_j} .
	\label{cranking_rotation}
\end{equation}

It is instructive to investigate properties of the cranking mass formula
for the NG modes \cite{RS80,BM75,TV62}.
Here, let us examine the formula of Eq. (\ref{cranking_translation})
for the translation, following the argument given in
Ref. \cite{BM75}.
The summation with respect to the index $n$ in Eq. (\ref{cranking_translation})
is restricted to the particle states.
This restriction can be removed, because
the summation with respect to the hole-hole components,
$\sum_{n\in h}\sum_{j\in h}$, identically vanishes.
When the mean-field (Hartree-Fock) Hamiltonian $\hat{h}_{\rm HF}[\rho]$
conserves the Galilean invariance,
\begin{equation}
	\left[ \hat{h}_{\rm HF}, \hat{x}_k \right] = \frac{-i}{m}\hat{p}_k,  \quad k=x,y,z ,
\end{equation}
the translational mass becomes
\begin{eqnarray}
	M_{\rm cr}^{\rm np} &=& 2\sum_{\mu} \sum_{j\in h}
	\frac{\bra{\varphi_j} \hat{p}_k \ket{\varphi_\mu}\bra{\varphi_\mu} \hat{p}_k \ket{\varphi_j}
	}{e_\mu-e_j} \nonumber\\
	&=& \sum_{\mu} \sum_{j\in h} \left(
	\bra{\varphi_j} im \left[ \hat{h}_{\rm HF}, \hat{x}_k \right]\ket{\varphi_\mu}\bra{\varphi_\mu} \hat{p}_k \ket{\varphi_j} \right. \nonumber\\
	&&\left.
	+\bra{\varphi_j} \hat{p}_k \ket{\varphi_\mu}\bra{\varphi_\mu} im \left[ \hat{h}_{\rm HF}, \hat{x}_k \right] \ket{\varphi_j} \right)
	\frac{1}{e_\mu-e_j} \nonumber\\
	&=& im \sum_{j\in h}
	\bra{\varphi_j} \left[\hat{p}_k, \hat{x}_k \right] \ket{\varphi_j}
	= Am .
	\label{cranking_total_mass}
\end{eqnarray}
Therefore, the Galilean invariance of the mean-field Hamiltonian guarantees
that the cranking formula reproduces the total mass $Am$ for the translation.
However, the Galilean invariance is violated by velocity dependence of the one-body
mean-field potential in most of nuclear energy density functionals.
This results in a wrong mass $M_{\rm cr}\neq Am$ in the cranking formula.
This violation should be corrected by the residual fields that depend on the velocity
of the translational motion \cite{BM75}.
Indeed, the ASCC mass reproduces the exact total mass even when the mean-field potential
violates the Galilean invariance, which will be shown in Sec.~\ref{sec:application}.

\section{\label{sec:application}Application}

In the following numerical calculations,
in order to express the orbital wave functions,
the grid representation is employed, discretizing the rectangular box into
three-dimensional (3D) Cartesian mesh \cite{NY05}.
The model space is set to be $12\times12\times18$ fm$^{3}$ for the
systems $\alpha$+$\alpha$ $\rightarrow$ $^{8}$Be
and $^{16}$O+$\alpha$ $\rightarrow$ $^{20}$Ne.
It is $12\times12\times24$ fm$^{3}$ for $^{16}$O+$^{16}$O $\rightarrow$ $^{32}$S.
The mesh size is set to be 1.0 fm for the system
$\alpha$+$\alpha$ $\rightarrow$ $^{8}$Be and 1.1 fm
for the other two systems.

For numerical calculations of the ASCC method,
we use the the finite amplitude method (FAM)~\citep{NIY07,INY09,AN11,AN13}.
The FAM provides us with high numerical efficiency with simple computer programs,
because only the calculations of the mean-field (single-particle) Hamiltonian
constructed with independent bra and ket states are required \cite{NIY07}.
The matrix FAM (m-FAM) prescription~\cite{AN13} is adopted
to solve the moving RPA equations (\ref{RPA0}) and (\ref{RPA}).
On the other hand, the iterative FAM (i-FAM)~\citep{NIY07,INY09,AN11}
is adopted to calculate the response functions for the NG modes.
The moving mean-field equation (\ref{chf}) is solved by using the
imaginary-time method \citep{DFKW80}.

\subsection{Modified BKN energy density functional}

In order to investigate the effect of this time-odd mean-field potential
on the collective inertial masses,
we adopt an energy density functional of the simplest choice,
namely, the BKN energy density functional \cite{BKN76} with the minimum extension.

The original BKN functional assumes the spin-isospin symmetry
without the spin-orbit interaction,
which is a functional of the isoscalar kinetic and local densities,
$\tau(\mathbf{r})$ and $\rho(\mathbf{r})$, only.
The mean-field potential is local and has no velocity dependence.
Thus, the nucleon's effective mass is identical to the bare nucleon mass.
However, in most of realistic energy density functionals,
the effective mass is smaller than the bare
mass, typically $m^{*}/m \sim 0.7$.
In order to introduce the effective mass, we extend the energy density by adding
terms $\rho\tau-\mathbf{j}^2$
where $\mathbf{j}(\mathbf{r})$ is the isoscalar
current density.
The appearance of the current density
is necessary to preserve
the Galilean invariance.

The modified BKN energy density functional reads,
\begin{eqnarray}
	E[\rho] &=& \int \frac{1}{2m} \tau(\mathbf{r}) d\mathbf{r}
	+\int d\mathbf{r} \left\{
            \frac{3}{8}t_{0}\rho^2(\mathbf{r})
           + \frac{1}{16}t_{3}\rho^{3}(\mathbf{r})\right\} \nonumber\\
   &&+\int \int d\mathbf{r} d\mathbf{r}' \rho(\mathbf{r})v(\mathbf{r}-\mathbf{r}')\rho(\mathbf{r}')\nonumber\\
	&&+B_{3} \int d\mathbf{r}
	\left\{ \rho(\mathbf{r})\tau(\mathbf{r}) - \mathbf{j}^{2}(\mathbf{r})\right\},
           \label{BKND}
\end{eqnarray}
where $\rho(\mathbf{r})$, $\tau(\mathbf{r})$, and $\mathbf{j}(\mathbf{r})$
are calculated as
\begin{eqnarray}
\rho(\mathbf{r})&=&4 \sum_{j=1}^{A/4}|\psi_{j}(\mathbf{r})|^{2}, \quad
	\tau(\mathbf{r})=4\sum_{j=1}^{A/4}|\nabla\psi_{j}(\mathbf{r})|^{2},\nonumber \\
	\textbf{j}(\mathbf{r})&=&\frac{4}{2i}\sum_{j=1}^{A/4}\left\{ \psi_{j}^{*}(\mathbf{r})\nabla\psi_{j}(\mathbf{r})
	-\psi_{j}(\mathbf{r})\nabla\psi_{j}^{*}(\mathbf{r}) \right\}.\nonumber
\end{eqnarray}
In Eq. (\ref{BKND}), $v(\mathbf{r})$ is the sum of the Yukawa and the Coulomb potentials \citep{BKN76},
\begin{eqnarray}
v(\mathbf{r})\equiv V_{0}a\frac{e^{-r/a}}{r} + \frac{(e/2)^{2}}{r}.\nonumber
\end{eqnarray}
The new parameter $B_3$ controls the effective mass and the velocity dependence of
the mean-field potential.

The variation of the total energy with respect to the density
(or equivalently single-particle wave functions)
defines the mean-field (Hartree-Fock) Hamiltonian,
\begin{eqnarray}
\hat{h}_{\rm HF}(\mathbf{r}) &=& -\nabla \frac{1}{2m^{*}(\mathbf{r})}\nabla
           + \frac{3}{4}t_{0}\rho(\mathbf{r})
           + \frac{3}{16}t_{3}\rho^{2}(\mathbf{r})\nonumber\\
        &&   + \int d\mathbf{r'} v(\mathbf{r}-\mathbf{r'})\rho(\mathbf{r'})
	   +B_{3}(\tau(\mathbf{r})+i\nabla\cdot \mathbf{j}(\mathbf{r}))\nonumber\\
	  &&  +2iB_{3}\mathbf{j}(\mathbf{r})\cdot\nabla,
	  \label{H_hf_BKN}
\end{eqnarray}
where the effective mass is now deviated from bare nucleon mass,
\begin{eqnarray}
	\frac{1}{2m^{*}(\mathbf{r})}=\frac{1}{2m}+B_{3}\rho(\mathbf{r}).
	\label{effective_mass}
\end{eqnarray}
For $B_3\neq 0$,
Eq. (\ref{H_hf_BKN})
indicates the velocity (momentum) dependence of the mean-field potential
and the presence of the time-odd mean fields,
$iB_3(\nabla\cdot\mathbf{j}+2\mathbf{j}\cdot\nabla)$.
For the time-even states, such as the ground state of even-even nuclei,
the current density disappears, $\mathbf{j}(\mathbf{r})=0$.
Nevertheless, as will be shown later, the terms associated with $\mathbf{j}(\mathbf{r})$ play
an important role in the collective inertial mass.

The parameters, $t_0$, $t_3$, $V_0$, and $a$
are adopted from Ref. \citep{BKN76},
and we vary $B_3$ to change the effective mass and the time-odd mean fields.

\subsection{Inertial massses for translational motion: Alpha particle}

First, we demonstrate the importance of the time-odd mean fields,
taking the translational total mass as a trivial example.
We adopt the simplest case, namely, the single alpha particle.
\begin{figure}
\begin{centering}
\includegraphics[width=0.90\columnwidth]{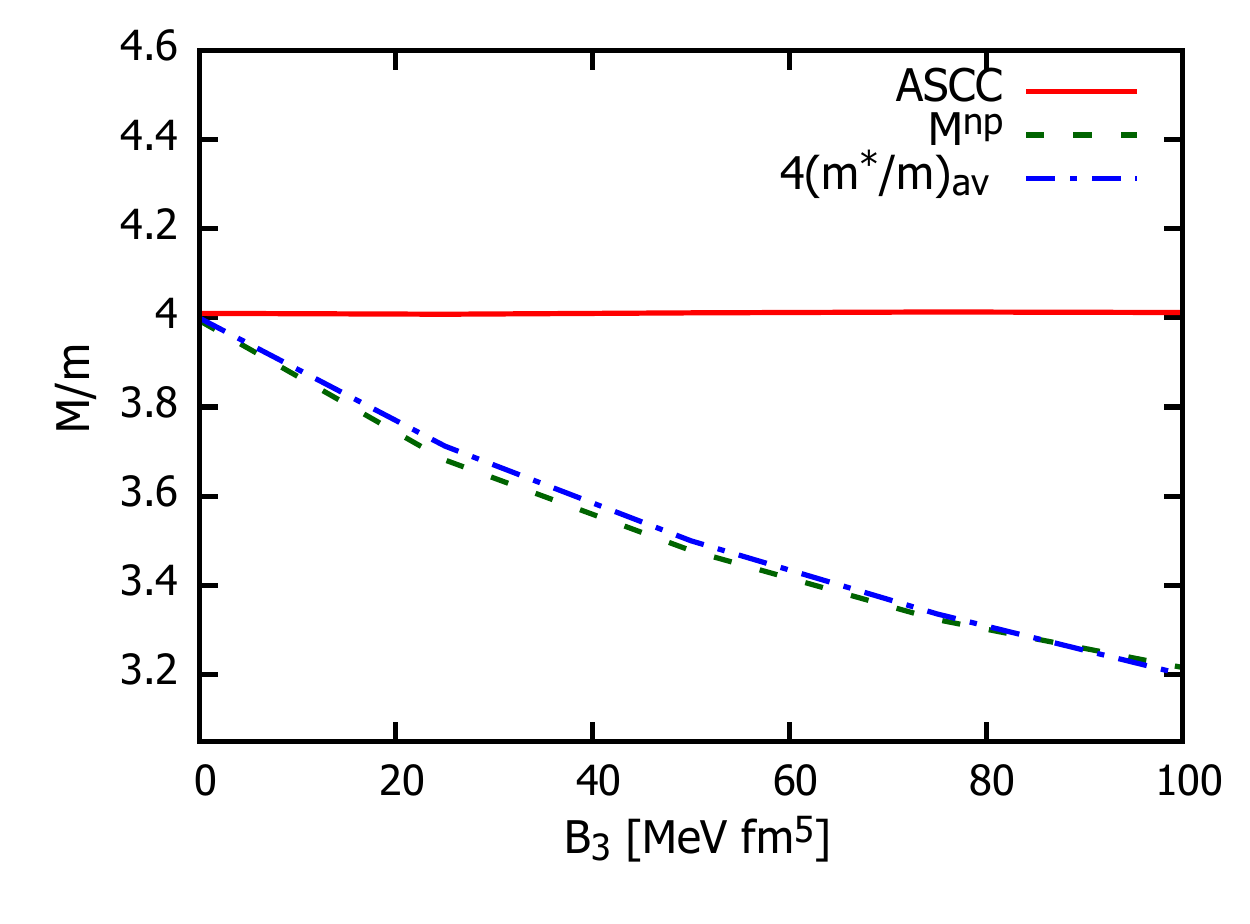}
\par\end{centering}
\caption{\label{fig:ma1}(Color online)
Translational mass of the $\alpha$ particle in unit of the nucleon mass as
a function of the parameter $B_{3}$.
The solid (red), dashed (green), and dash-dotted (blue) curves
show the results
of the ASCC, the non-perturbative cranking formula
(\ref{cranking_translation})
and $A$ times the average effective mass, respectively.
See the text for details.
}
\end{figure}
In Fig. \ref{fig:ma1}, we show the translational mass of a single alpha particle
as a function of $B_{3}$. 
Increasing $B_3$, the effective mass $m^*$ decreases,
as we see in Eq. (\ref{effective_mass}).
In the present case, we use the center-of-mass coordinate $\mathbf{R}_{\rm cm}$
as the collective coordinate $R$
in Sec.~\ref{sec:theoa1} and \ref{sec:CHF+cranking}.
Since the system is isotropic, we use its $z$ component $Z_{\rm cm}$ in
the numerical calculation.
In the present model, neutrons and protons have the identical mass $m$.
Therefore, the total inertial mass of an alpha particle should be equal to $4m$.

It is clearly demonstrated that
the ASCC always reproduces
the correct total mass $Am$, irrespective of values of the parameter $B_3$.
The non-perturbative cranking mass reproduces the total mass, $M^{\rm np}_{\rm cr}=4m$,
at $B_3=0$, which agree with Eq. (\ref{cranking_total_mass}).
This is due to the velocity-independent mean-field potential, for which
the mean-field Hamiltonian $\hat{h}_{\rm HF}$ preserves the Galilean invariance.
However, at $B_3>0$, the effective mass becomes smaller than
the bare nucleon mass, $m^* < m$, inside the nucleus with $\rho(\mathbf{r})>0$.
This leads to the velocity-dependent potential, and the violation of
the Galilean invariance
in the mean-field level.
Figure~\ref{fig:ma1} indicates $M_{\rm cr}^{\rm np}<4m$ at $B_3>0$,
and shows a monotonic decrease with respect to $B_3$.
We calculate the average effective mass defined by
\begin{equation}
	\left(\frac{m^*}{m}\right)_{\rm av}^{-1} \equiv \frac{2m}{A}
	\int \frac{\rho(\mathbf{r})}{2m^*(\mathbf{r})} d\mathbf{r} .
	\label{average_effective_mass}
\end{equation}
It turns out that the cranking mass for the translation
approximately coincides with
$Am\times (m^*/m)_{\rm av}$ as shown in
Fig. \ref{fig:ma1}.


For the translational motion,
the ASCC inertial mass becomes identical to the Thouless-Valatin
moments of inertia \cite{TV62} in Eqs. (\ref{RPA0-2}) and (\ref{RPA-2}),
with $(\hat{J}_x,\hat{\Theta})$ being replaced by $(\hat{P}_z,\hat{Z}_{\rm cm})$.
The fact that the ASCC inertial mass agrees with the total mass $Am$
means that the Galilean invariance, violated
in the mean-field level,
is properly restored by the residual induced fields in the leading order.
This shows a clear advantageous feature of the ASCC mass over the cranking
formulae.


\subsection{\label{sec:appa}Inertial massses for relative motions }

\subsubsection{\label{sec:appa1}$\alpha$+$\alpha$ $\rightarrow ^{8}$Be }

\begin{figure}
\begin{centering}
\includegraphics[width=0.90\columnwidth]{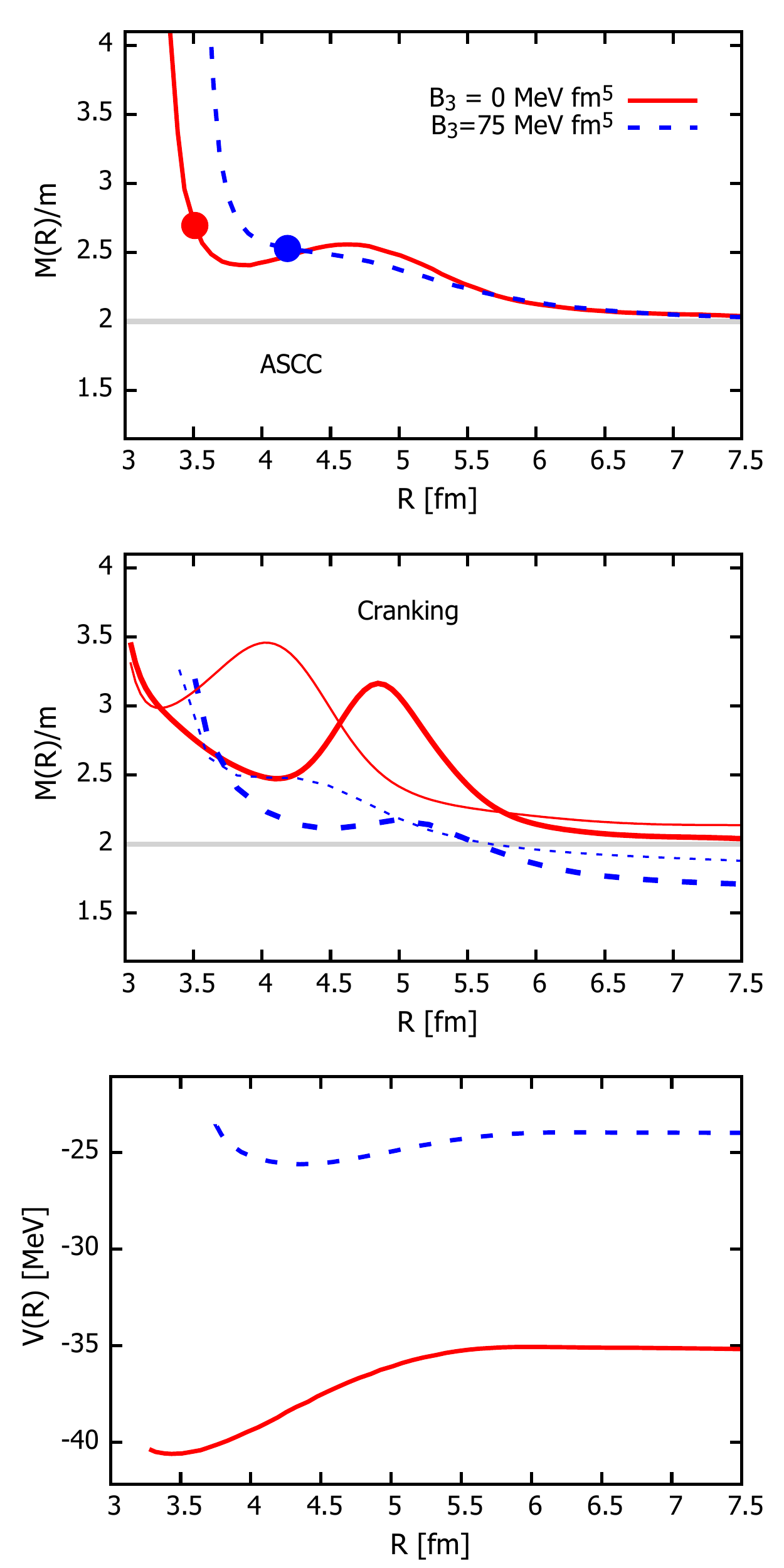}
\par\end{centering}
\caption{\label{fig:aa2}(Color online)
Inertial mass $M(R)$ for the reaction $\alpha+\alpha$
as a function of the relative distance $R$.
The top panel shows the results of the ASCC method, while
the middle panel shows those of the perturbative and non-perturbative
cranking formulae $M_{\rm cr}^{\rm p}$ (thin lines)
and $M_{\rm cr}^{\rm np}$ (thick lines).
The bottom panel shows the potential energies
along the ASCC collective paths.
Solid (red) and dashed (blue) lines
indicate those of $B_{3}=0$ and 75  MeV fm$^5$, respectively.
In the upper panel, the dots correspond to the positions of the
minimum energy in the bottom panel.
}
\end{figure}

In this section we study how the time-odd mean-field changes the inertial
mass for the relative motion between two alpha particles.
The reaction path was obtained and shown in our former paper \cite{WenN16} for $B_3=0$.
With $B_3\neq 0$, basic features of the reaction path is the same,
although the ground state of $^8$Be becomes more elongated, developing
a prominent $2\alpha$ cluster structure.
The curves in Fig. \ref{fig:aa2} show the inertial mass and total energy for the relative
motion between two alpha particles
as a function of the relative distance $R$.
The ASCC mass is calculated based on the
self-consistent ASCC collective path.
For the cranking masses, the mass quadrupole operator $\hat{Q}_{20}$
is used as a constraint to construct the reaction path with the CHF calculation.

\begin{figure}
\begin{centering}
\includegraphics[width=0.90\columnwidth]{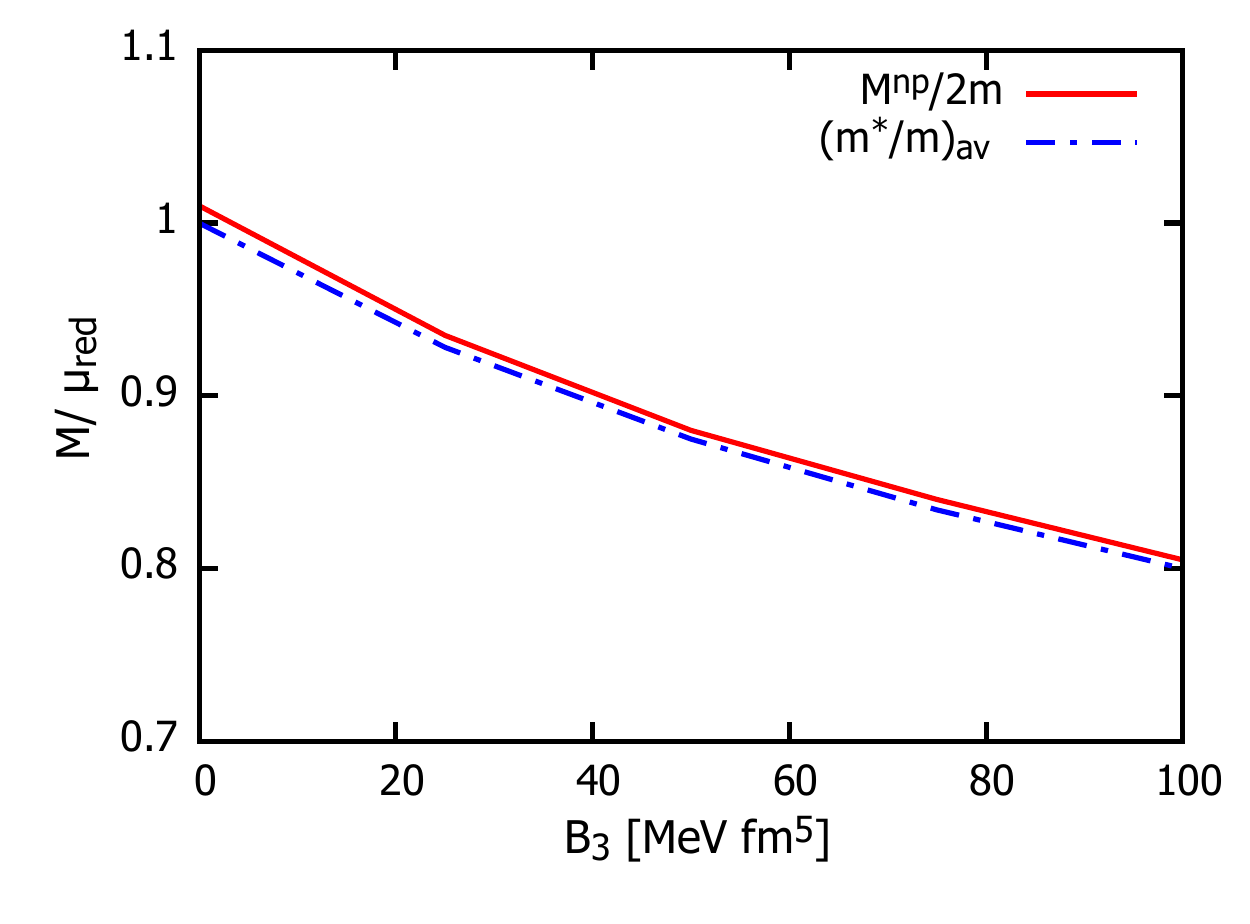}
\par\end{centering}
\caption{\label{fig:a1a2}(Color online)
Non-perturbative cranking inertial masses
calculated at $R=8$ fm
for the relative motion between two $\alpha$ particles
divided by the reduced mass $\mu_{\rm red}=2m$,
as a function of $B_{3}$. 
The blue dash-dotted curve indicates $(m^*/m)_{\rm av}$
of Eq. (\ref{average_effective_mass}).
}
\end{figure}

At large $R$ where the two alpha particles can be
approximated as point particles,
we expect that the inertial mass with respect to $R$ is identical to
the reduced mass, $2m$.
For the non-perturbative cranking mass, this is true at $B_3=0$,
while it monotonically decreases as $B_3$ increases
(Fig.~\ref{fig:a1a2}).
On the other hand, the ASCC mass reproduces the correct
reduced mass at large $R$, irrespective of the value of $B_3$.
The main difference between the ASCC and the cranking masses is
the inclusion of the effect of time-odd residual fields.
Therefore, we may conclude that the time-odd residual effect is essential
to reproduce the reduced mass in the asymptotic region of $R\rightarrow\infty$.


The perturbative cranking mass seems
to be unable
to reproduce the correct reduced mass.
It is larger than $\mu_{\rm red}=2m$ even for $B_3=0$.
As is seen in the first equation of (\ref{cranking_approximation}),
the non-perturbative cranking mass comes from
the $R$-dependence of the Lagrange multiplier $\lambda$ and
from that of the HF Hamiltonian (rearrangement) $\hat{h}_{\rm HF}$.
When we take the asymptotic limit of $R\rightarrow\infty$,
since the interaction between two $\alpha$ particles
vanishes, we naturally expect $\lambda\rightarrow 0$.
Then, we have $d\lambda/dR\rightarrow 0$,
and $M^{\rm np}_{\rm cr}$
solely comes from the rearrangement effect.
In the perturbative treatment, we neglect the rearrangement
in the last equality of Eq. (\ref{cranking_approximation}).
Instead, we replace $d\lambda/dR$ by the perturbative value
$(2S_1(\hat{R},\hat{C}))^{-1}$.
It is thus
difficult to justify the perturbative approximation
in the asymptotic limit.

Figure~\ref{fig:a1a2} clearly demonstrates
that the cranking inertial mass at $B_3\neq 0$
underestimates the exact reduced mass
for the relative motion ($\mu_{\rm red}=2m$).
Again, it well agrees
with the average effective mass,
$2(m^*)_{\rm av}$.
Thus, the failure of the cranking mass is due to
the violation of the Galilean invariance.


\subsubsection{\label{sec:theoa2}
$\alpha$+$^{16}$O $\rightarrow ^{20}$Ne and $^{16}$O+$^{16}$O $\rightarrow ^{32}$S  }

\begin{figure}
\begin{centering}
\includegraphics[width=0.90\columnwidth]{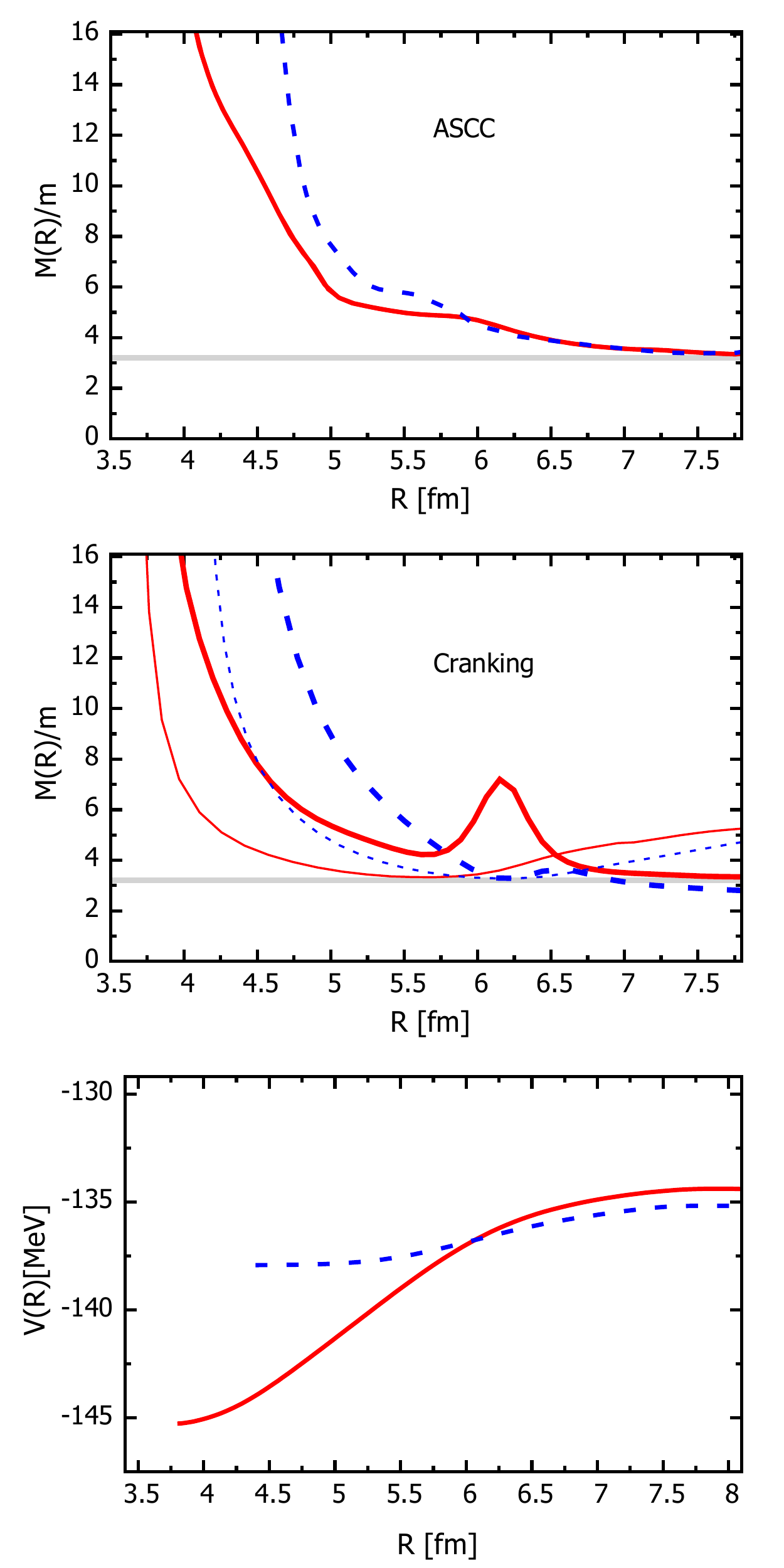}
\par\end{centering}
\caption{\label{fig:aOr}(Color online)
Inertial mass $M(R)$ for the reaction path $\alpha$+$^{16}$O
as a function
of $R$, in units of the nucleon mass.
The top panel shows the results of the ASCC method,
the middle panel shows those of the perturbative and non-perturbative
cranking formulae $M_{\rm cr}^{\rm p}$ (thin lines) and $M_{\rm cr}^{\rm np}$
(thick lines),
and the bottom panel shows the potential energies along the three ASCC
collective paths.
Solid (red) and dashed (blue) lines
indicate those of $B_{3}=0$ and 75  MeV fm$^5$.
In the bottom panel,
in order to show it in the given scale,
the potential for $B_3=75$ MeV fm$^5$ (dashed line)
is shifted downwards by 60 MeV.
}
\end{figure}

\begin{figure}
\begin{centering}
\includegraphics[width=0.90\columnwidth]{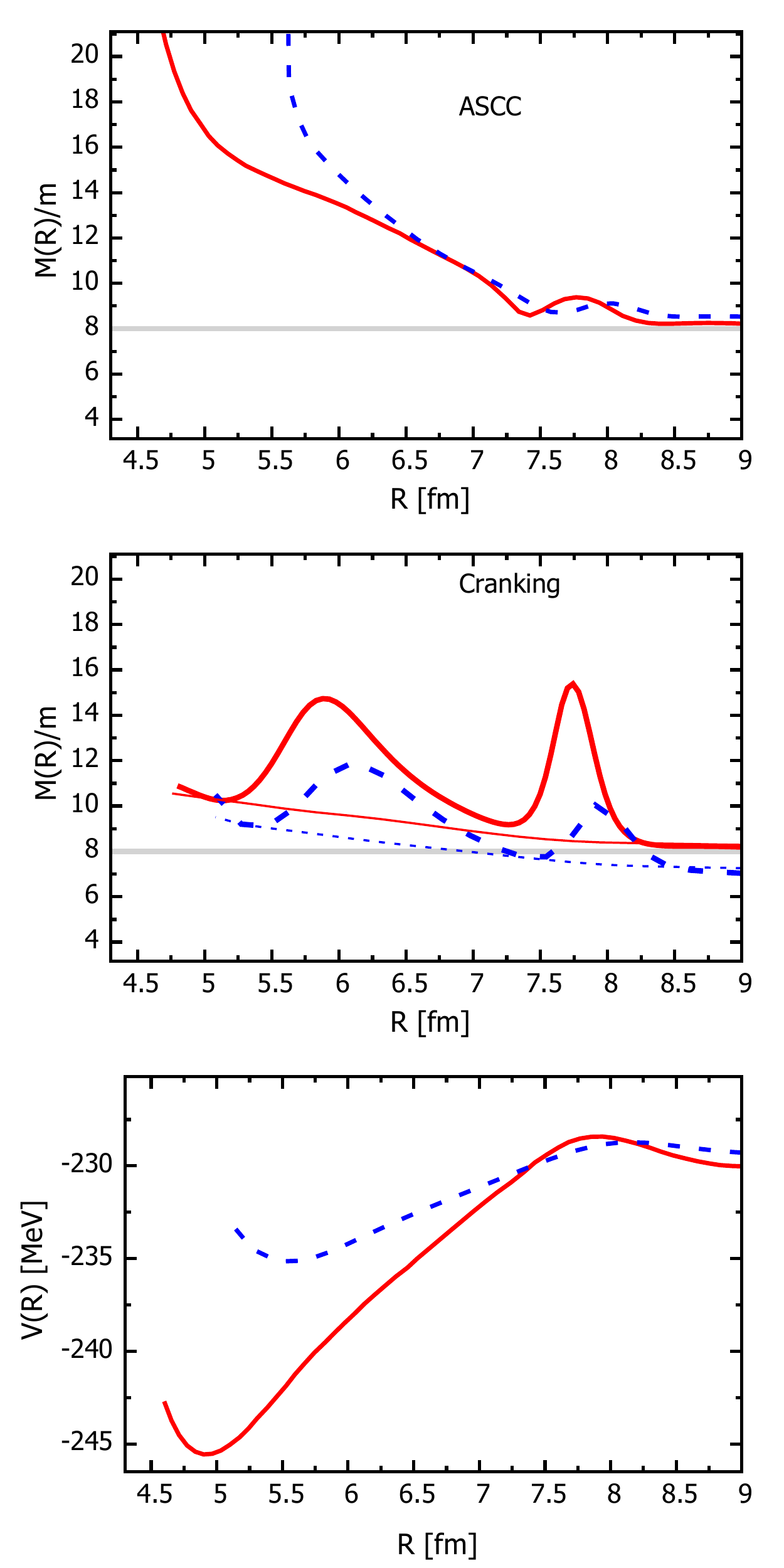}
\par\end{centering}
\caption{\label{fig:OOr}(Color online)
Same as Fig.~\ref{fig:aOr}, but for $^{16}$O+$^{16}$O.
Solid (red) and dashed (blue) lines
indicate results of $B_{3}=0$ and 30  MeV fm$^5$.
In the bottom panel,
the potential for $B_3=30$ MeV fm$^5$ (dashed line)
is shifted downwards by 45 MeV.
}
\end{figure}

Next we show the inertial masses $M(R)$ of the relative motion
for two reaction systems $\alpha$+$^{16}$O $\rightarrow ^{20}$Ne
and $^{16}$O+$^{16}$O $\rightarrow ^{32}$S.
The self-consistent reaction paths for these were presented
in our former paper \cite{WenN17}, but only for $B_3=0$.
Similar to the $\alpha+\alpha\rightarrow^8$Be case,
the increase of $B_3$ favors larger deformation.
The upper panels in Figs.~\ref{fig:aOr} and \ref{fig:OOr} show
$M(R)$ for $\alpha$+$^{16}$O $\rightarrow$ $^{20}$Ne
and $^{16}$O+$^{16}$O $\rightarrow$ $^{32}$S, respectively.
The smaller effective mass enlarges the size of the nucleus,
thus, the touching points between the two nuclei take place at larger $R$.
The minimum energy position is also shifted to larger $R$
for larger $B_3$.
The drastic increase in $M(R)$ in the interior region is due to
the increase of $dq/dR$ in Eq. (\ref{mass}).
In contrast to variations of $M(R)$ in the interior region,
$M(R)$ is very stable and independent from the value of $B_3$
when two nuclei are separated.
In the asymptotic region ($R\rightarrow\infty$),
we always observe the correct limit, $M(R)\rightarrow\mu_{\rm red}$.
The reduced mass for $\alpha$+$^{16}$O $\rightarrow ^{20}$Ne is
$\mu_{\rm red}= 3.2 m$, and $\mu_{\rm red}= 8 m$
for $^{16}$O+$^{16}$O $\rightarrow ^{32}$S.

The cranking masses are shown in the middle panels
in Figs. \ref{fig:aOr} and \ref{fig:OOr}.
The perturbative and non-perturbative cranking masses
are calculated based on the CHF reaction paths
that are constructed
with the constraining operator $\hat{C}=\hat{Q}_{30}$ (mass octupole)
for $\alpha$+$^{16}$O $\rightarrow^{20}$Ne,
and with $\hat{C}=\hat{Q}_{20}$ (mass quadrupole)
for $^{16}$O$+^{16}$O $\rightarrow^{32}$S.
If we adopt $\hat{C}=\hat{Q}_{20}$ for the former reaction system,
we cannot obtain a continuous reaction path \cite{WenN17}.
Again, for $B_3\neq 0$,
the cranking mass does not reproduce the correct reduced mass
at $R\rightarrow\infty$,
neither with the perturbative cranking formula
nor with the non-perturbative one.
Furthermore, the perturbative cranking mass is larger than
the reduced mass, even for $B_3=0$.

\subsection{\label{sec:rotation}Rotational moments of inertia}

Next let us present results for the moment of inertia which is
another important inertial mass for the collective Hamiltonian in Eq. (\ref{reaction_model}).
The rotational motion is a NG mode whose ASCC inertial mass can be calculated
with the strength function for the angular momentum operator
at zero frequency \cite{Hin15},
instead of solving the Thouless-Valatin equation (\ref{RPA-2}).
For this purpose, we perform the i-FAM calculation \cite{NIY07,INY09,AN11}
based on the reaction path self-consistently constructed by the ASCC method.
For comparison,
we also apply the cranking formula to the moments of inertia,
Eq. (\ref{cranking_rotation}), based on the reaction path
calculated with the CHF method.

In Eq. (\ref{reaction_model}), the point-particle approximation was adopted,
leading to the moment of inertia, $\mu_{\rm red}R^2$.
There is another classical limit, namely,
the rigid-body moment of inertia.
In the present case, the system preserves
the axial symmetry on the reaction path.
Choosing the axis of the symmetry as the $z$ axis,
we have ${\cal J}_x = {\cal J}_y$ whose rigid-body value is given by
\begin{eqnarray}
{\cal J}_{\rm rig}=m\int (y^{2}+z^{2})\rho(\mathbf{r})d\mathbf{r}
=m\int (z^{2}+x^{2})\rho(\mathbf{r})d\mathbf{r}.
\end{eqnarray}
The rigid-body moment of inertia
about the $z$ axis is non-zero, ${\cal J}_z\neq 0$,
which contradicts with the trivial quantum mechanical requirement that
there exists no rotation around the symmetry axis.
On the other hand,
it is known that the cranking moments of inertia,
${\cal J}_x$ and ${\cal J}_y$,
for axially deformed nuclei
in the harmonic-oscillator-potential model are given
by the rigid-body value at the equilibrium \cite{BM75}.
The cranking formula also satisfies the quantum mechanical condition,
producing ${\cal J}_z=0$ in the axially symmetric case.

When two nuclei are far away,
we expect the point-particle approximation is good.
Therefore, we expect that the moments of inertia,
${\cal J}_x$ and ${\cal J}_y$, on the reaction path
changes from the point-particle value ${\cal J}(R)=\mu_{\rm red}R^2$
at large $R$
to the rigid-body value ${\cal J}(R)={\cal J}_{\rm rig}$
near the equilibrium (ground) state.
It is of significant interest and of importance to examine
where and how this transition takes place
during the reaction.

\begin{figure}
\begin{centering}
\includegraphics[width=0.90\columnwidth]{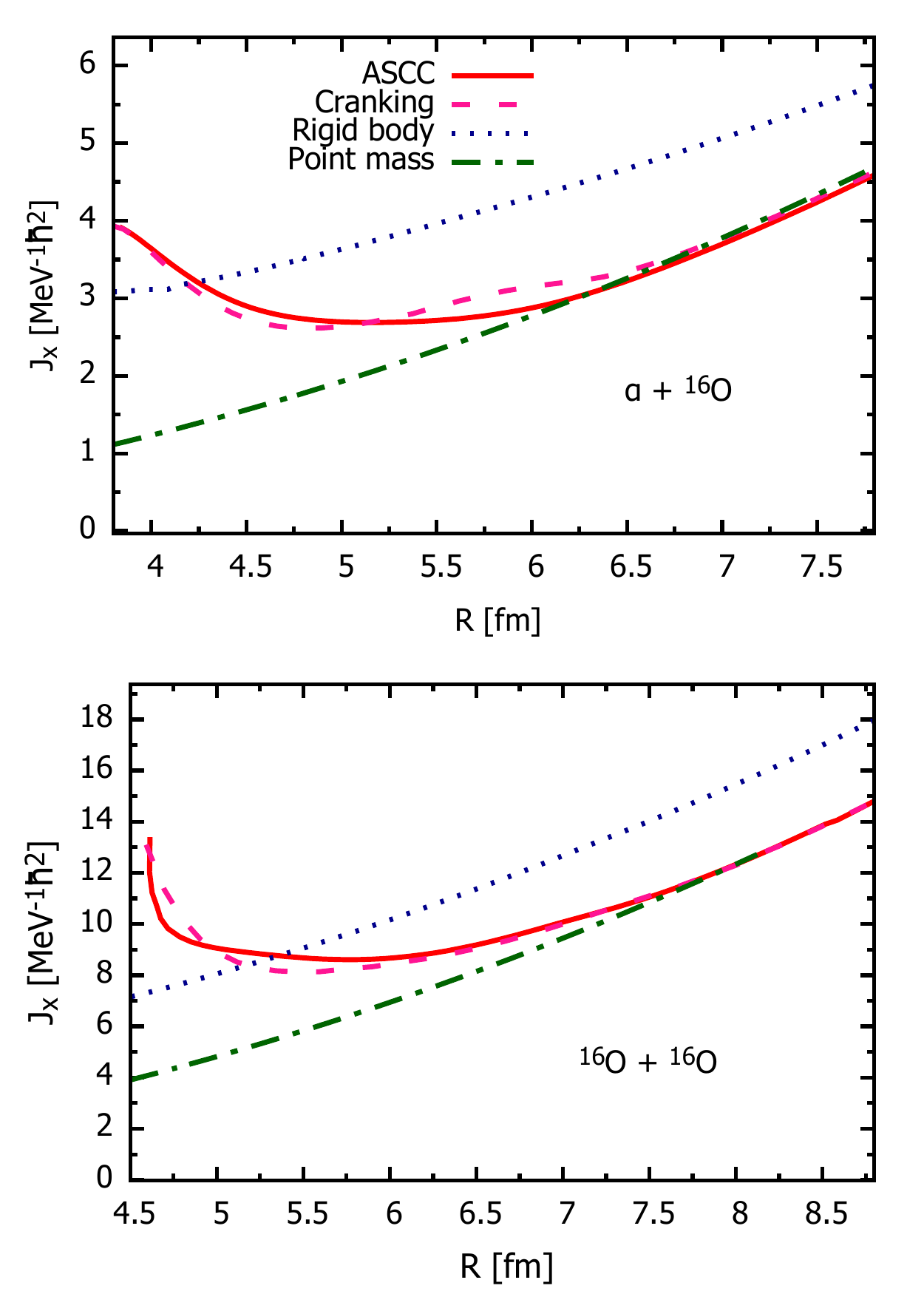}
\par\end{centering}
\caption{\label{fig:jxnes}(Color online)
Calculated rotational moments of inertias for $B_3=0$
as a function of relative distance $R$.
The upper panel shows the results for the system $\alpha$+$^{16}$O,
the lower panel shows the results for the system $^{16}$O+$^{16}$O.
The solid (red), dashed (pink), dotted (blue) and dash-dotted (green)
curves indicate the results of ASCC, non-perturbative cranking formula,
rigid body approximation, and point-mass approximation, respectively.
}
\end{figure}

In Ref.~\cite{WenN20}, we have published the result
for $\alpha+\alpha\rightarrow^8$Be.
Thus, in this paper, we present the results for
$\alpha$+$^{16}$O $\rightarrow ^{20}$Ne and
$^{16}$O+$^{16}$O $\rightarrow ^{32}$S.
First, let us show results for the velocity-independent potential with $B_3=0$.
Figure~\ref{fig:jxnes} shows
the moments of inertia calculated
for the two reaction systems with $B_3=0$.
The ground state of $^{20}$Ne corresponds to
$R = 3.8$ fm in the upper panel of Fig.~\ref{fig:jxnes},
and the superdeformed minimum of $^{32}$S is located at $R = 5$ fm
in the lower panel.
We can see that the ASCC and the (non-perturbative) cranking formulae
give results very similar to each other.
This is the consequence of the local Galilean invariance
of the mean-field potential under the momentum transformation
$\mathbf{p}\rightarrow\mathbf{p}-m(\boldsymbol{\omega}\times\mathbf{r})$ \cite{BM75}.
The calculated moments of inertia are close to their rigid-body values
near these equilibrium states,
${\cal J}(R_e)\approx {\cal J}_{\rm rig}(R_e)$,
where $R_e$ represents the value of $R$ at the potential equilibrium
($dV/dR=0$).
However, when the nucleus is more elongated along the reaction path,
${\cal J}(R)$ decreases as $R$ increases.
Since the rigid-body value is a monotonically increasing function of $R$,
they become smaller than the rigid-body values, ${\cal J}(R) <{\cal J}_{\rm rig}(R)$,
then, quickly approach the point-particle values, $\mu_{\rm red}R^2$.
Beyond the scission point $R=R_s$ where the system splits into two separated nuclei, we have
${\cal J}(R)\approx \mu_{\rm red}R^2$ ($R\geq R_s$).
The calculated moments of inertia give parabolic convex lines as functions of $R$,
showing their minima located around the midpoint between $R_e$ and $R_s$.

It is surprising that the moments of inertia decrease as
the deformation develops near the equilibrium.
This contradicts with our naive expectation based on the classical model.
For instance,
the rigid-body moments of inertia linearly increases
with the deformation $\delta$,
${\cal J}_{\rm rig}\approx {\cal J}_0 (1+\delta/3)$.
Another classical model, irrotational fluid model suggests
that it increases quadratically with the deformation as
${\cal J}_{\rm irrot}\approx {\cal J}_{\rm rig}\delta^2$ \cite{BM75}.
In any case, a single classical model cannot explain the reduction
in ${\cal J}(R)$ as a function of $R$.
A hint to understand this behavior may come
from the harmonic oscillator model in the quantum mechanics
which reproduces both the rigid-body
value and the irrotational-fluid value depending on
the configuration of nucleus \cite{BM75}.
During the evolution of deformation as a function of $R$,
different configurations appear, which may lead to moments of inertia
corresponding to different classical models.
This suggests the importance of the quantum mechanical calculation.

Another striking feature in Fig.~\ref{fig:jxnes} is
${\cal J}(R)\approx \mu_{\rm red}R^2$ at $R\geq R_s$,
which is significantly smaller than ${\cal J}_{\rm rig}(R)$.
When the two nuclei are separated, the structure of
the projectile and the target nuclei are almost invariant with respect to $R$.
Let us denote the rigid-body moments of inertia of the projectile and the target
with respect to their own center-of-mass coordinates as
${\cal J}_P$ and ${\cal J}_T$, respectively.
The total rigid-body moments of inertia can be written as
\begin{equation}
{\cal J}(R)={\cal J}_P + {\cal J}_T + \mu_{\rm red} R^2 .
	\label{MoI_PTR}
\end{equation}
This equation is easy to prove in the rigid-body case.
In the quantum mechanical treatment of the present cases,
there is another trivial result,
namely, ${\cal J}_P = {\cal J}_T =0$.
This is because
both the projectile and the target nuclei ($\alpha$ and $^{16}$O)
are spherical in the ground state,
thus, the quantum mechanics requires the vanishing moments of inertia
(See also Eq. (\ref{cranking_rotation})).
Assuming that Eq.~(\ref{MoI_PTR}) is valid for the quantum mechanical systems,
we find ${\cal J}(R)\approx \mu_{\rm red}R^2$
in the region where the projectile and the target become spherical.
This is also a consequence of the quantum mechanics.

\begin{figure}
\begin{centering}
\includegraphics[width=0.90\columnwidth]{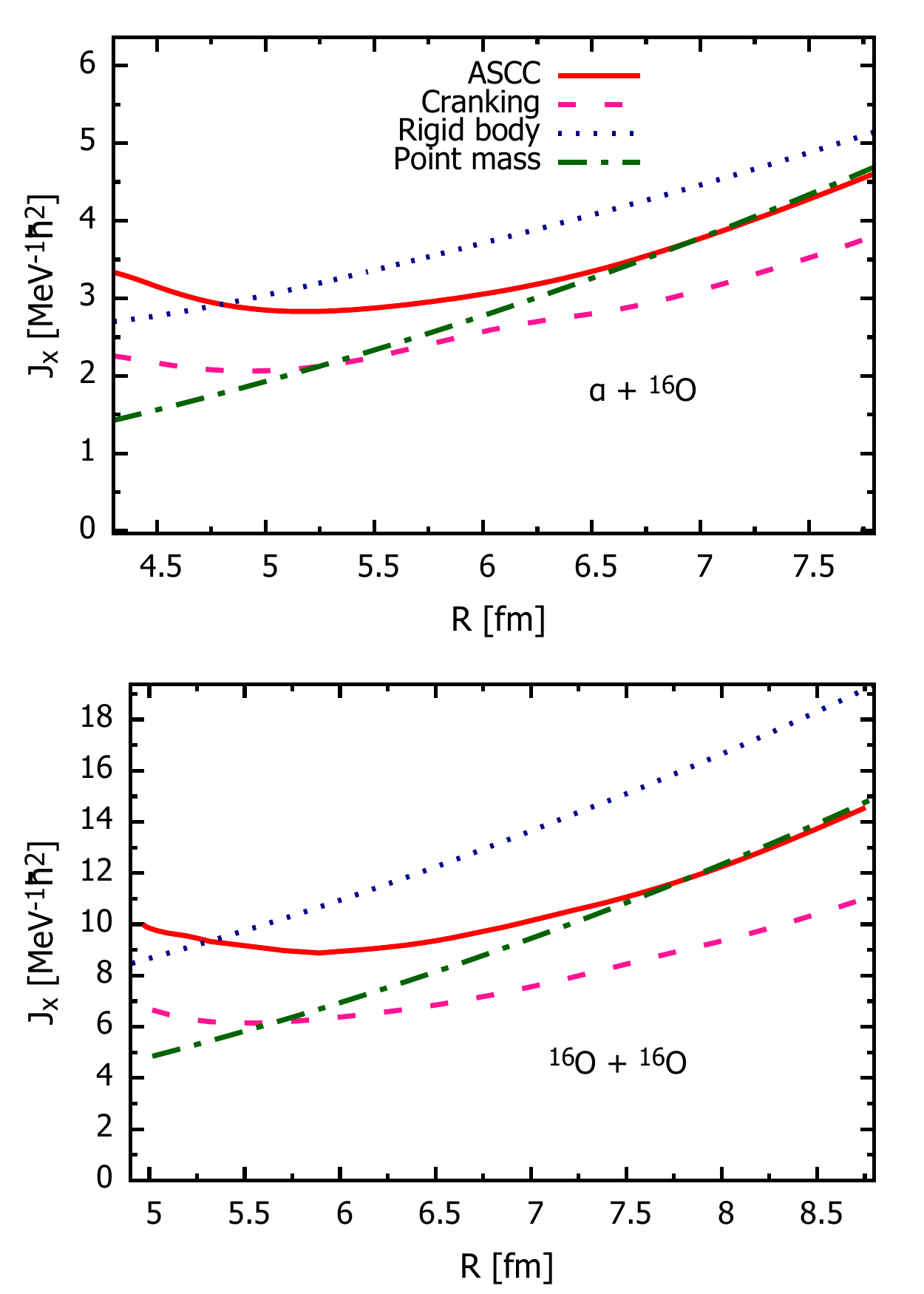}
\par\end{centering}
\caption{\label{fig:rotb3}(Color online)
Same as Fig.~\ref{fig:jxnes}, but for $B_3=25$ MeV fm$^{5}$.
}
\end{figure}

Next, let us show how the $B_{3}$ term influences the moments of inertia.
Figure~\ref{fig:rotb3} shows the results.
For $B_{3}\neq 0$, the velocity dependence in the mean-field potential violates the
local Galilean invariance.
The calculated cranking moments of inertia are significantly smaller than
the rigid-body value ${\cal J}_{\rm rig}$ at $R\approx R_e$.
They are also smaller than
the point-particle value $\mu_{\rm red}R^2$ at $R>R_s$.
On the other hand, the ASCC calculation includes the residual effects of
the time-odd mean fields which restore the local Galilean invariance,
then, nicely reproduces both rigid-body and point-particle values
at $R\approx R_e$ and at $R>R_s$, respectively.
In fact, for these reaction systems,
the ASCC moments of inertia are insensitive to the $B_3$ value
over the entire region.


\subsection{\label{sec:theoc}Impact on astrophysical $S$ factors}

\begin{figure}
\begin{centering}
\includegraphics[width=0.90\columnwidth]{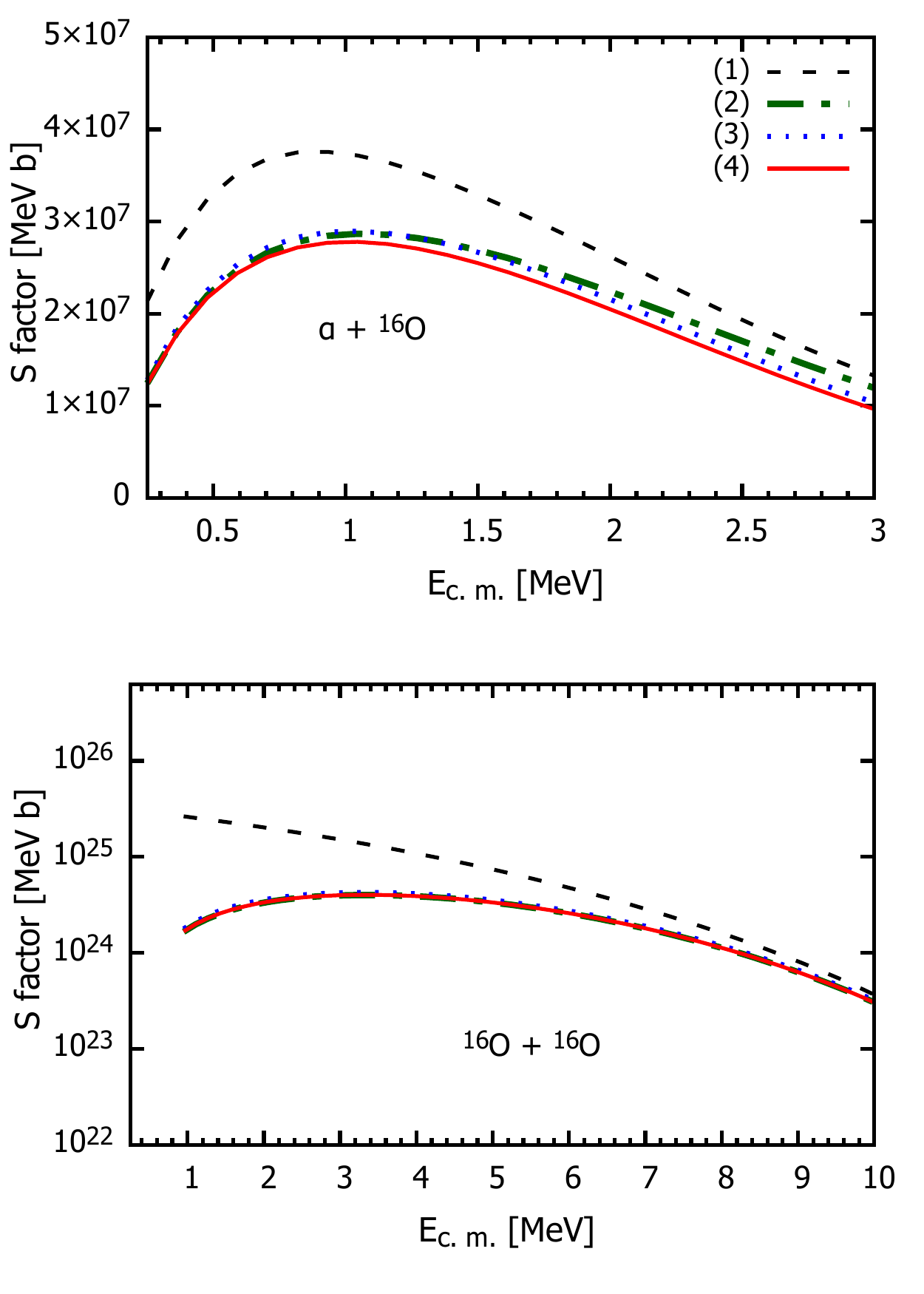}
\par\end{centering}
\caption{\label{fig:sf}(Color online)
Astrophysical $S$ factor calculated with $B_3=0$ for
$\alpha+^{16}$O (upper panel) and for $^{16}$O+$^{16}$O (lower panel),
as a function of incident energy $E_{\rm c.m.}$.
(1) The point-particle approximation
both for the relative and the rotational degrees of freedom,
$\mu^{\rm red}$ and $\mu_{\rm red}R^2$, respectively.
(2) The ASCC mass $M(R)$ for the relative motion together with
$\mu_{\rm red}R^2$ for the rotation.
(3) ASCC $M(R)$ and rigid-body ${\cal J}_{\rm rig}(R)$.
(4) ASCC $M(R)$ and ASCC ${\cal J}(R)$.
See text for details.
}
\end{figure}

\begin{figure}
\begin{centering}
\includegraphics[width=0.90\columnwidth]{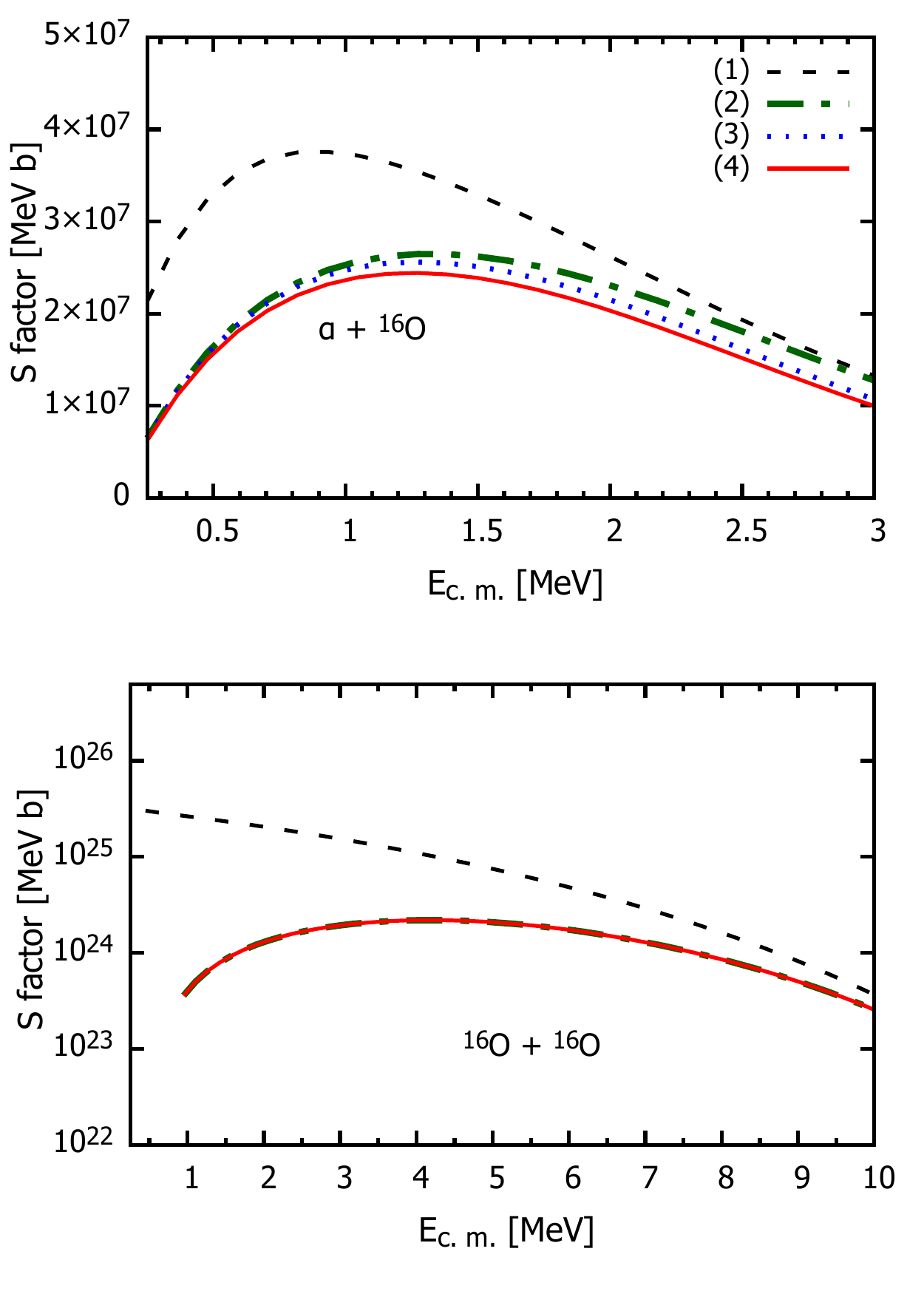}
\par\end{centering}
\caption{\label{fig:sf2}(Color online)
Same as Fig.~\ref{fig:sf}, but calculated with the inertial masses
for $B_3=25$ MeV fm$^5$.
The same potentials as those of $B_3=0$ are used.
See text for details.
}
\end{figure}

In this section, we present the calculation of the astrophysical $S$ factor
for fusion reactions, using the potentials and the inertial masses obtained
in Sec.~\ref{sec:appa} and \ref{sec:rotation}.
The ASCC calculation provides us with the collective Hamiltonian
with the optimal reaction paths for $^{16}$O+$\alpha$ $\rightarrow$ $^{20}$Ne and
$^{16}$O+$^{16}$O$\rightarrow ^{32}$S.
The Hamiltonian for the reaction is given as that of Eq. (\ref{reaction_model})
in which the moment of inertia $\mu_{\rm red}R^2$ is replaced by ${\cal J}(R)$,
and the reduced mass $\mu_{\rm red}$ by $M(R)$.
We investigate the dependence of
the sub-barrier fusion cross sections on these
inertial masses.

The sub-barrier fusion cross section is calculated with the WKB approximation,
following the procedure in Ref.~\cite{RFGGG84}.
The total sub-barrier fusion cross section is given by the sum of
all the partial waves with angular moment $\ell$.
Under the WKB approximation,
the transmission coefficient for the partial wave $\ell$
at incident energy $E_{\rm c.m.}$ is given by
\begin{eqnarray}
T_{\ell}(E_{\rm c.m.}) = [1+\exp(2I_{\ell})]^{-1},
\end{eqnarray}
with
\begin{eqnarray}
I_{\ell}(E_{\rm c.m.})&=& \int_{a}^{b}dR \Big\{ 2M(R) \nonumber \\
	&\times& \left( V(R)
	+\frac{\ell(\ell+1)}{2{\cal J}(R)} - E_{\rm c.m.} \right) \Big\}^{1/2} ,
	 \label{I_L}
\end{eqnarray}
where $a$ and $b$ are the classical turning points on the inner and outer
sides of the barrier respectively.
In addition to the potential $V(R)$,
the coordinate-dependent inertial mass for the relative
motion $M(R)$ and for the rotational moment of inertia ${\cal J}(R)$ are
necessary for the calculation.
The term $\ell(\ell+1)/2{\cal J}(R)$ represents the
centrifugal potential.

The fusion cross section is given by
\begin{eqnarray}
	\sigma(E_{\rm c.m.}) = \frac{\pi}{2\mu_{\rm red} E_{\rm c.m.}}\sum_{\ell}(2\ell+1)T_{\ell}(E_{\rm c.m.}) .
	\label{sigma_fus}
\end{eqnarray}
For identical incident nuclei, $^{16}$O+$^{16}$O,
Eq. (\ref{sigma_fus}) must be modified according to the proper symmetrization.
Only the partial wave with even $L$ contribute to the cross section as
\begin{eqnarray}
	\sigma(E_{\rm c.m.}) = \frac{\pi}{2\mu_{\rm red} E_{\rm c.m.}}\sum_{\ell}[1+(-)^{\ell}](2\ell+1)T_{\ell}(E_{\rm c.m.}).\nonumber \\
\end{eqnarray}
Instead of $\sigma(E_{\rm c.m.})$, we show the astrophysical $S$ factor defined by
\begin{eqnarray}
S(E_{\rm c.m.}) = E_{\rm c.m.}\sigma(E_{\rm c.m.}) \exp[ 2\pi Z_{1}Z_{2}e^{2}/\hbar v ],
\end{eqnarray}
where $v$ is the relative velocity at $R\rightarrow\infty$.
The astrophysical $S$ factor is preferred for sub-barrier fusion
because it removes the change by tens of orders of magnitude
present in the cross section due to the penetration through the Coulomb barrier.
The $S$ factor can reveal in a more transparent way
the influence of the nuclear structure and dynamics.

Figure \ref{fig:sf} shows the calculated $S$ factor with $B_3=0$ for the
scattering of $\alpha+^{16}$O (upper panel) and
$^{16}$O+$^{16}$O (lower panel), respectively.
In order to clarify the effect of the inertial masses,
we use the same potential $V(R)$ for all the curves in each panel
of Fig.~\ref{fig:sf},
which is the ASCC potential obtained for $B_3=0$
(solid lines in the bottom panels of Fig.~\ref{fig:aOr}
and \ref{fig:OOr}).
Different curves show the calculations with different
inertia masses.
For the case of $B_3=0$, the ASCC and the cranking inertial masses are
similar to each other.
Thus, the calculation with the cranking inertial masses produces
the astrophysical $S$ factor similar to the ASCC result.
Generally speaking,
the larger the inertial mass $M(R)$ for the relative motion is,
the smaller the astrophysical $S$ factor is.
The opposite effect can be seen for the rotational moments of inertia ${\cal J}(R)$.
The larger moments of inertia give the larger $S$ factor.
The replacement of the constant mass $\mu_{\rm red}$ by ASCC $M(R)$
gives a significant effect for the fusion cross section.
For the reaction system $^{16}$O+$^{16}$O, we find
a strong suppression of the fusion cross section at $E_{\rm c.m.} < 4$ MeV
compared to the calculation with $\mu_{\rm red}$,
which is an order of magnitude or even larger.

Figure \ref{fig:sf2} shows the same $S$ factors with
inertial masses calculated with $B_3=25$ MeV fm$^5$ for the
scattering of $\alpha+^{16}$O (upper panel) and
$^{16}$O+$^{16}$O (lower panel), respectively.
In order to see the effect of inertial masses,
we use the potential $V(R)$ obtained for $B_3=0$.
Thus, the differences between Figs.~\ref{fig:sf} and \ref{fig:sf2}
come from the change in inertial masses.
Again, the change from the reduced mass $\mu_{\rm red}$ into $M(R)$
gives the largest impact on the $S$ factor.
The suppression effect of $M(R)$ is even more significant than
the $B_3=0$ case, because the $M(R)$ is larger for finite values of $B_3$,
as we can find in Fig.~\ref{fig:aOr} and \ref{fig:OOr}.

Although the BKN density functional provides us only with qualitative results,
these calculations reveal the important roles of the inertial masses,
and suggest significant influence of the inertial masses
on the sub-barrier fusion cross sections.

\section{\label{sec:sum}Summary}

We calculated the ASCC inertial mass coefficients
with respect to the translational, relative and rotational motions.
The numerical calculations are performed using the FAM
in the 3D coordinate space representation for the reaction systems
$\alpha$+$\alpha$ $\rightarrow$ $^{8}$Be,
$\alpha+^{16}$O $\rightarrow$ $^{20}$Ne
and $^{16}$O$+^{16}$O $\rightarrow$ $^{32}$S.
We investigated the impact of time-odd mean-field potentials on
the collective inertial masses.
In the presence of nucleonic effective mass,
the cranking mass can neither
reproduce the total mass $Am$ of translational motion
nor the correct reduced mass $\mu_{\rm red}$ for the relative motion.
In addition,
the cranking formula for the rotation produces
neither
the point-particle value $\mu_{\rm red}R^2$ in the asymptotic region
($R\rightarrow\infty$),
nor the rigid-body value ${\cal J}_{\rm rid}$ near the equilibrium states.
In contrast,
the ASCC masses is able to properly take into account
the residual time-odd effects, producing the total mass $Am$
for translation,
the reduced mass $\mu_{\rm red}$ for relative motion,
and the point-particle value $\mu_{\rm red}R^2$
(the rigid-body value ${\cal J}_{\rm rid}$)
after the scission (near the equilibrium).
We also calculated the astrophysical $S$ factors
with
these microscopic inputs as the inertial masses of a reaction model.
It turns out that replacement of the reduced mass $\mu_{\rm red}$
by $M(R)$ obtained with ASCC gives the largest impact on the
$S$ factors.
It could lead to a strong suppression of the fusion cross section
at low energy.

In the present study,
a schematic BKN interaction plus time-odd terms are adopted for a
qualitative investigation.
It is highly desired to use
realistic modern nuclear energy functionals.
The paring correlation is also expected to play a critical role
in low-energy collective dynamics.
These issues are currently under investigation.

\begin{acknowledgments}
This work is supported in part by JSPS KAKENHI Grant No. 18H01209,
No. 19H05142, and No. 20K14458.
This research in part used computational resources provided through the HPCI System Research
Project (Project ID: hp200069),
and by Multidisciplinary Cooperative Research Program in Center for
Computational Sciences, University of Tsukuba.
\end{acknowledgments}

\bibliography{myself,nuclear_physics,current,current2}

\begin{thebibliography}{43}%
\makeatletter
\providecommand \@ifxundefined [1]{%
 \@ifx{#1\undefined}
}%
\providecommand \@ifnum [1]{%
 \ifnum #1\expandafter \@firstoftwo
 \else \expandafter \@secondoftwo
 \fi
}%
\providecommand \@ifx [1]{%
 \ifx #1\expandafter \@firstoftwo
 \else \expandafter \@secondoftwo
 \fi
}%
\providecommand \natexlab [1]{#1}%
\providecommand \enquote  [1]{``#1''}%
\providecommand \bibnamefont  [1]{#1}%
\providecommand \bibfnamefont [1]{#1}%
\providecommand \citenamefont [1]{#1}%
\providecommand \href@noop [0]{\@secondoftwo}%
\providecommand \href [0]{\begingroup \@sanitize@url \@href}%
\providecommand \@href[1]{\@@startlink{#1}\@@href}%
\providecommand \@@href[1]{\endgroup#1\@@endlink}%
\providecommand \@sanitize@url [0]{\catcode `\\12\catcode `\$12\catcode
  `\&12\catcode `\#12\catcode `\^12\catcode `\_12\catcode `\%12\relax}%
\providecommand \@@startlink[1]{}%
\providecommand \@@endlink[0]{}%
\providecommand \url  [0]{\begingroup\@sanitize@url \@url }%
\providecommand \@url [1]{\endgroup\@href {#1}{\urlprefix }}%
\providecommand \urlprefix  [0]{URL }%
\providecommand \Eprint [0]{\href }%
\providecommand \doibase [0]{http://dx.doi.org/}%
\providecommand \selectlanguage [0]{\@gobble}%
\providecommand \bibinfo  [0]{\@secondoftwo}%
\providecommand \bibfield  [0]{\@secondoftwo}%
\providecommand \translation [1]{[#1]}%
\providecommand \BibitemOpen [0]{}%
\providecommand \bibitemStop [0]{}%
\providecommand \bibitemNoStop [0]{.\EOS\space}%
\providecommand \EOS [0]{\spacefactor3000\relax}%
\providecommand \BibitemShut  [1]{\csname bibitem#1\endcsname}%
\let\auto@bib@innerbib\@empty
\bibitem [{\citenamefont {Negele}(1982)}]{Neg82}%
  \BibitemOpen
  \bibfield  {author} {\bibinfo {author} {\bibfnamefont {J.~W.}\ \bibnamefont
  {Negele}},\ }\href {\doibase 10.1103/RevModPhys.54.913} {\bibfield  {journal}
  {\bibinfo  {journal} {Rev. Mod. Phys.}\ }\textbf {\bibinfo {volume} {54}},\
  \bibinfo {pages} {913} (\bibinfo {year} {1982})}\BibitemShut {NoStop}%
\bibitem [{\citenamefont {Simenel}(2012)}]{Sim12}%
  \BibitemOpen
  \bibfield  {author} {\bibinfo {author} {\bibfnamefont {C.}~\bibnamefont
  {Simenel}},\ }\href {\doibase 10.1140/epja/i2012-12152-0} {\bibfield
  {journal} {\bibinfo  {journal} {The European Physical Journal A}\ }\textbf
  {\bibinfo {volume} {48}},\ \bibinfo {pages} {152} (\bibinfo {year}
  {2012})}\BibitemShut {NoStop}%
\bibitem [{\citenamefont {Maruhn}\ \emph {et~al.}(2014)\citenamefont {Maruhn},
  \citenamefont {Reinhard}, \citenamefont {Stevenson},\ and\ \citenamefont
  {Umar}}]{MRSU14}%
  \BibitemOpen
  \bibfield  {author} {\bibinfo {author} {\bibfnamefont {J.~A.}\ \bibnamefont
  {Maruhn}}, \bibinfo {author} {\bibfnamefont {P.-G.}\ \bibnamefont
  {Reinhard}}, \bibinfo {author} {\bibfnamefont {P.~D.}\ \bibnamefont
  {Stevenson}}, \ and\ \bibinfo {author} {\bibfnamefont {A.~S.}\ \bibnamefont
  {Umar}},\ }\href {\doibase http://dx.doi.org/10.1016/j.cpc.2014.04.008}
  {\bibfield  {journal} {\bibinfo  {journal} {Computer Physics Communications}\
  }\textbf {\bibinfo {volume} {185}},\ \bibinfo {pages} {2195 } (\bibinfo
  {year} {2014})}\BibitemShut {NoStop}%
\bibitem [{\citenamefont {Nakatsukasa}\ and\ \citenamefont
  {Yabana}(2005)}]{NY05}%
  \BibitemOpen
  \bibfield  {author} {\bibinfo {author} {\bibfnamefont {T.}~\bibnamefont
  {Nakatsukasa}}\ and\ \bibinfo {author} {\bibfnamefont {K.}~\bibnamefont
  {Yabana}},\ }\href@noop {} {\bibfield  {journal} {\bibinfo  {journal} {Phys.
  Rev. C}\ }\textbf {\bibinfo {volume} {71}},\ \bibinfo {eid} {024301}
  (\bibinfo {year} {2005})}\BibitemShut {NoStop}%
\bibitem [{\citenamefont {Nakatsukasa}\ \emph {et~al.}(2016)\citenamefont
  {Nakatsukasa}, \citenamefont {Matsuyanagi}, \citenamefont {Matsuo},\ and\
  \citenamefont {Yabana}}]{NMMY16}%
  \BibitemOpen
  \bibfield  {author} {\bibinfo {author} {\bibfnamefont {T.}~\bibnamefont
  {Nakatsukasa}}, \bibinfo {author} {\bibfnamefont {K.}~\bibnamefont
  {Matsuyanagi}}, \bibinfo {author} {\bibfnamefont {M.}~\bibnamefont {Matsuo}},
  \ and\ \bibinfo {author} {\bibfnamefont {K.}~\bibnamefont {Yabana}},\ }\href
  {\doibase 10.1103/RevModPhys.88.045004} {\bibfield  {journal} {\bibinfo
  {journal} {Rev. Mod. Phys.}\ }\textbf {\bibinfo {volume} {88}},\ \bibinfo
  {pages} {045004} (\bibinfo {year} {2016})}\BibitemShut {NoStop}%
\bibitem [{\citenamefont {Sekizawa}\ and\ \citenamefont {Yabana}(2016)}]{SY16}%
  \BibitemOpen
  \bibfield  {author} {\bibinfo {author} {\bibfnamefont {K.}~\bibnamefont
  {Sekizawa}}\ and\ \bibinfo {author} {\bibfnamefont {K.}~\bibnamefont
  {Yabana}},\ }\href {\doibase 10.1103/PhysRevC.93.054616} {\bibfield
  {journal} {\bibinfo  {journal} {Phys. Rev. C}\ }\textbf {\bibinfo {volume}
  {93}},\ \bibinfo {pages} {054616} (\bibinfo {year} {2016})}\BibitemShut
  {NoStop}%
\bibitem [{\citenamefont {Umar}\ \emph {et~al.}(2016)\citenamefont {Umar},
  \citenamefont {Oberacker},\ and\ \citenamefont {Simenel}}]{UOS16}%
  \BibitemOpen
  \bibfield  {author} {\bibinfo {author} {\bibfnamefont {A.~S.}\ \bibnamefont
  {Umar}}, \bibinfo {author} {\bibfnamefont {V.~E.}\ \bibnamefont {Oberacker}},
  \ and\ \bibinfo {author} {\bibfnamefont {C.}~\bibnamefont {Simenel}},\ }\href
  {\doibase 10.1103/PhysRevC.94.024605} {\bibfield  {journal} {\bibinfo
  {journal} {Phys. Rev. C}\ }\textbf {\bibinfo {volume} {94}},\ \bibinfo
  {pages} {024605} (\bibinfo {year} {2016})}\BibitemShut {NoStop}%
\bibitem [{\citenamefont {Sekizawa}(2017)}]{Sek17}%
  \BibitemOpen
  \bibfield  {author} {\bibinfo {author} {\bibfnamefont {K.}~\bibnamefont
  {Sekizawa}},\ }\href {\doibase 10.1103/PhysRevC.96.014615} {\bibfield
  {journal} {\bibinfo  {journal} {Phys. Rev. C}\ }\textbf {\bibinfo {volume}
  {96}},\ \bibinfo {pages} {014615} (\bibinfo {year} {2017})}\BibitemShut
  {NoStop}%
\bibitem [{\citenamefont {Umar}\ \emph {et~al.}(2017)\citenamefont {Umar},
  \citenamefont {Simenel},\ and\ \citenamefont {Ye}}]{USY17}%
  \BibitemOpen
  \bibfield  {author} {\bibinfo {author} {\bibfnamefont {A.~S.}\ \bibnamefont
  {Umar}}, \bibinfo {author} {\bibfnamefont {C.}~\bibnamefont {Simenel}}, \
  and\ \bibinfo {author} {\bibfnamefont {W.}~\bibnamefont {Ye}},\ }\href
  {\doibase 10.1103/PhysRevC.96.024625} {\bibfield  {journal} {\bibinfo
  {journal} {Phys. Rev. C}\ }\textbf {\bibinfo {volume} {96}},\ \bibinfo
  {pages} {024625} (\bibinfo {year} {2017})}\BibitemShut {NoStop}%
\bibitem [{\citenamefont {Ebata}\ \emph {et~al.}(2010)\citenamefont {Ebata},
  \citenamefont {Nakatsukasa}, \citenamefont {Inakura}, \citenamefont
  {Yoshida}, \citenamefont {Hashimoto},\ and\ \citenamefont {Yabana}}]{Eba10}%
  \BibitemOpen
  \bibfield  {author} {\bibinfo {author} {\bibfnamefont {S.}~\bibnamefont
  {Ebata}}, \bibinfo {author} {\bibfnamefont {T.}~\bibnamefont {Nakatsukasa}},
  \bibinfo {author} {\bibfnamefont {T.}~\bibnamefont {Inakura}}, \bibinfo
  {author} {\bibfnamefont {K.}~\bibnamefont {Yoshida}}, \bibinfo {author}
  {\bibfnamefont {Y.}~\bibnamefont {Hashimoto}}, \ and\ \bibinfo {author}
  {\bibfnamefont {K.}~\bibnamefont {Yabana}},\ }\href {\doibase
  10.1103/PhysRevC.82.034306} {\bibfield  {journal} {\bibinfo  {journal} {Phys.
  Rev. C}\ }\textbf {\bibinfo {volume} {82}},\ \bibinfo {pages} {034306}
  (\bibinfo {year} {2010})}\BibitemShut {NoStop}%
\bibitem [{\citenamefont {Stetcu}\ \emph {et~al.}(2011)\citenamefont {Stetcu},
  \citenamefont {Bulgac}, \citenamefont {Magierski},\ and\ \citenamefont
  {Roche}}]{SBMR11}%
  \BibitemOpen
  \bibfield  {author} {\bibinfo {author} {\bibfnamefont {I.}~\bibnamefont
  {Stetcu}}, \bibinfo {author} {\bibfnamefont {A.}~\bibnamefont {Bulgac}},
  \bibinfo {author} {\bibfnamefont {P.}~\bibnamefont {Magierski}}, \ and\
  \bibinfo {author} {\bibfnamefont {K.~J.}\ \bibnamefont {Roche}},\ }\href
  {\doibase 10.1103/PhysRevC.84.051309} {\bibfield  {journal} {\bibinfo
  {journal} {Phys. Rev. C}\ }\textbf {\bibinfo {volume} {84}},\ \bibinfo
  {pages} {051309} (\bibinfo {year} {2011})}\BibitemShut {NoStop}%
\bibitem [{\citenamefont {Hashimoto}(2012)}]{Has12}%
  \BibitemOpen
  \bibfield  {author} {\bibinfo {author} {\bibfnamefont {Y.}~\bibnamefont
  {Hashimoto}},\ }\href {\doibase 10.1140/epja/i2012-12055-0} {\bibfield
  {journal} {\bibinfo  {journal} {The European Physical Journal A}\ }\textbf
  {\bibinfo {volume} {48}},\ \bibinfo {pages} {55} (\bibinfo {year}
  {2012})}\BibitemShut {NoStop}%
\bibitem [{\citenamefont {Hashimoto}(2013)}]{Has13}%
  \BibitemOpen
  \bibfield  {author} {\bibinfo {author} {\bibfnamefont {Y.}~\bibnamefont
  {Hashimoto}},\ }\href {\doibase 10.1103/PhysRevC.88.034307} {\bibfield
  {journal} {\bibinfo  {journal} {Phys. Rev. C}\ }\textbf {\bibinfo {volume}
  {88}},\ \bibinfo {pages} {034307} (\bibinfo {year} {2013})}\BibitemShut
  {NoStop}%
\bibitem [{\citenamefont {Hashimoto}\ and\ \citenamefont
  {Scamps}(2016)}]{HS16}%
  \BibitemOpen
  \bibfield  {author} {\bibinfo {author} {\bibfnamefont {Y.}~\bibnamefont
  {Hashimoto}}\ and\ \bibinfo {author} {\bibfnamefont {G.}~\bibnamefont
  {Scamps}},\ }\href {\doibase 10.1103/PhysRevC.94.014610} {\bibfield
  {journal} {\bibinfo  {journal} {Phys. Rev. C}\ }\textbf {\bibinfo {volume}
  {94}},\ \bibinfo {pages} {014610} (\bibinfo {year} {2016})}\BibitemShut
  {NoStop}%
\bibitem [{\citenamefont {Bulgac}\ \emph {et~al.}(2016)\citenamefont {Bulgac},
  \citenamefont {Magierski}, \citenamefont {Roche},\ and\ \citenamefont
  {Stetcu}}]{BMRS16}%
  \BibitemOpen
  \bibfield  {author} {\bibinfo {author} {\bibfnamefont {A.}~\bibnamefont
  {Bulgac}}, \bibinfo {author} {\bibfnamefont {P.}~\bibnamefont {Magierski}},
  \bibinfo {author} {\bibfnamefont {K.~J.}\ \bibnamefont {Roche}}, \ and\
  \bibinfo {author} {\bibfnamefont {I.}~\bibnamefont {Stetcu}},\ }\href
  {\doibase 10.1103/PhysRevLett.116.122504} {\bibfield  {journal} {\bibinfo
  {journal} {Phys. Rev. Lett.}\ }\textbf {\bibinfo {volume} {116}},\ \bibinfo
  {pages} {122504} (\bibinfo {year} {2016})}\BibitemShut {NoStop}%
\bibitem [{\citenamefont {Magierski}\ \emph {et~al.}(2017)\citenamefont
  {Magierski}, \citenamefont {Sekizawa},\ and\ \citenamefont
  {Wlaz\l{}owski}}]{MSW17}%
  \BibitemOpen
  \bibfield  {author} {\bibinfo {author} {\bibfnamefont {P.}~\bibnamefont
  {Magierski}}, \bibinfo {author} {\bibfnamefont {K.}~\bibnamefont {Sekizawa}},
  \ and\ \bibinfo {author} {\bibfnamefont {G.}~\bibnamefont {Wlaz\l{}owski}},\
  }\href {\doibase 10.1103/PhysRevLett.119.042501} {\bibfield  {journal}
  {\bibinfo  {journal} {Phys. Rev. Lett.}\ }\textbf {\bibinfo {volume} {119}},\
  \bibinfo {pages} {042501} (\bibinfo {year} {2017})}\BibitemShut {NoStop}%
\bibitem [{\citenamefont {Scamps}\ and\ \citenamefont
  {Hashimoto}(2019)}]{SH19}%
  \BibitemOpen
  \bibfield  {author} {\bibinfo {author} {\bibfnamefont {G.}~\bibnamefont
  {Scamps}}\ and\ \bibinfo {author} {\bibfnamefont {Y.}~\bibnamefont
  {Hashimoto}},\ }\href {\doibase 10.1103/PhysRevC.100.024623} {\bibfield
  {journal} {\bibinfo  {journal} {Phys. Rev. C}\ }\textbf {\bibinfo {volume}
  {100}},\ \bibinfo {pages} {024623} (\bibinfo {year} {2019})}\BibitemShut
  {NoStop}%
\bibitem [{\citenamefont {Ring}\ and\ \citenamefont {Schuck}(1980)}]{RS80}%
  \BibitemOpen
  \bibfield  {author} {\bibinfo {author} {\bibfnamefont {P.}~\bibnamefont
  {Ring}}\ and\ \bibinfo {author} {\bibfnamefont {P.}~\bibnamefont {Schuck}},\
  }\href@noop {} {\emph {\bibinfo {title} {The nuclear many-body problems}}},\
  Texts and monographs in physics\ (\bibinfo  {publisher} {Springer-Verlag},\
  \bibinfo {address} {New York},\ \bibinfo {year} {1980})\BibitemShut {NoStop}%
\bibitem [{\citenamefont {Matsuo}\ \emph {et~al.}(2000)\citenamefont {Matsuo},
  \citenamefont {Nakatsukasa},\ and\ \citenamefont {Matsuyanagi}}]{MNM00}%
  \BibitemOpen
  \bibfield  {author} {\bibinfo {author} {\bibfnamefont {M.}~\bibnamefont
  {Matsuo}}, \bibinfo {author} {\bibfnamefont {T.}~\bibnamefont {Nakatsukasa}},
  \ and\ \bibinfo {author} {\bibfnamefont {K.}~\bibnamefont {Matsuyanagi}},\
  }\href@noop {} {\bibfield  {journal} {\bibinfo  {journal} {Prog. Theor.
  Phys.}\ }\textbf {\bibinfo {volume} {103}},\ \bibinfo {pages} {959} (\bibinfo
  {year} {2000})}\BibitemShut {NoStop}%
\bibitem [{\citenamefont {Hinohara}\ \emph {et~al.}(2007)\citenamefont
  {Hinohara}, \citenamefont {Nakatsukasa}, \citenamefont {Matsuo},\ and\
  \citenamefont {Matsuyanagi}}]{HNMM07}%
  \BibitemOpen
  \bibfield  {author} {\bibinfo {author} {\bibfnamefont {N.}~\bibnamefont
  {Hinohara}}, \bibinfo {author} {\bibfnamefont {T.}~\bibnamefont
  {Nakatsukasa}}, \bibinfo {author} {\bibfnamefont {M.}~\bibnamefont {Matsuo}},
  \ and\ \bibinfo {author} {\bibfnamefont {K.}~\bibnamefont {Matsuyanagi}},\
  }\href@noop {} {\bibfield  {journal} {\bibinfo  {journal} {Prog. Theor.
  Phys.}\ }\textbf {\bibinfo {volume} {117}},\ \bibinfo {pages} {451} (\bibinfo
  {year} {2007})}\BibitemShut {NoStop}%
\bibitem [{\citenamefont {Hinohara}\ \emph {et~al.}(2009)\citenamefont
  {Hinohara}, \citenamefont {Nakatsukasa}, \citenamefont {Matsuo},\ and\
  \citenamefont {Matsuyanagi}}]{HNMM09}%
  \BibitemOpen
  \bibfield  {author} {\bibinfo {author} {\bibfnamefont {N.}~\bibnamefont
  {Hinohara}}, \bibinfo {author} {\bibfnamefont {T.}~\bibnamefont
  {Nakatsukasa}}, \bibinfo {author} {\bibfnamefont {M.}~\bibnamefont {Matsuo}},
  \ and\ \bibinfo {author} {\bibfnamefont {K.}~\bibnamefont {Matsuyanagi}},\
  }\href {\doibase 10.1103/PhysRevC.80.014305} {\bibfield  {journal} {\bibinfo
  {journal} {Phys. Rev. C}\ }\textbf {\bibinfo {volume} {80}},\ \bibinfo
  {pages} {014305} (\bibinfo {year} {2009})}\BibitemShut {NoStop}%
\bibitem [{\citenamefont {Nakatsukasa}(2012)}]{Nak12}%
  \BibitemOpen
  \bibfield  {author} {\bibinfo {author} {\bibfnamefont {T.}~\bibnamefont
  {Nakatsukasa}},\ }\href {\doibase 10.1093/ptep/pts016} {\bibfield  {journal}
  {\bibinfo  {journal} {Progress of Theoretical and Experimental Physics}\
  }\textbf {\bibinfo {volume} {2012}},\ \bibinfo {pages} {01A207} (\bibinfo
  {year} {2012})}\BibitemShut {NoStop}%
\bibitem [{\citenamefont {Marumori}\ \emph {et~al.}(1980)\citenamefont
  {Marumori}, \citenamefont {Maskawa}, \citenamefont {Sakata},\ and\
  \citenamefont {Kuriyama}}]{MMSK80}%
  \BibitemOpen
  \bibfield  {author} {\bibinfo {author} {\bibfnamefont {T.}~\bibnamefont
  {Marumori}}, \bibinfo {author} {\bibfnamefont {T.}~\bibnamefont {Maskawa}},
  \bibinfo {author} {\bibfnamefont {F.}~\bibnamefont {Sakata}}, \ and\ \bibinfo
  {author} {\bibfnamefont {A.}~\bibnamefont {Kuriyama}},\ }\href {\doibase
  10.1143/PTP.64.1294} {\bibfield  {journal} {\bibinfo  {journal} {Progress of
  Theoretical Physics}\ }\textbf {\bibinfo {volume} {64}},\ \bibinfo {pages}
  {1294} (\bibinfo {year} {1980})}\BibitemShut {NoStop}%
\bibitem [{\citenamefont {Klein}\ \emph {et~al.}(1991)\citenamefont {Klein},
  \citenamefont {Walet},\ and\ \citenamefont {{Do Dang}}}]{KWD91}%
  \BibitemOpen
  \bibfield  {author} {\bibinfo {author} {\bibfnamefont {A.}~\bibnamefont
  {Klein}}, \bibinfo {author} {\bibfnamefont {N.~R.}\ \bibnamefont {Walet}}, \
  and\ \bibinfo {author} {\bibfnamefont {G.}~\bibnamefont {{Do Dang}}},\ }\href
  {\doibase https://doi.org/10.1016/0003-4916(91)90343-7} {\bibfield  {journal}
  {\bibinfo  {journal} {Annals of Physics}\ }\textbf {\bibinfo {volume}
  {208}},\ \bibinfo {pages} {90} (\bibinfo {year} {1991})}\BibitemShut
  {NoStop}%
\bibitem [{\citenamefont {Dang}\ \emph {et~al.}(2000)\citenamefont {Dang},
  \citenamefont {Klein},\ and\ \citenamefont {Walet}}]{DKW00}%
  \BibitemOpen
  \bibfield  {author} {\bibinfo {author} {\bibfnamefont {G.~D.}\ \bibnamefont
  {Dang}}, \bibinfo {author} {\bibfnamefont {A.}~\bibnamefont {Klein}}, \ and\
  \bibinfo {author} {\bibfnamefont {N.~R.}\ \bibnamefont {Walet}},\ }\href
  {\doibase 10.1016/S0370-1573(99)00119-2} {\bibfield  {journal} {\bibinfo
  {journal} {Physics Reports}\ }\textbf {\bibinfo {volume} {335}},\ \bibinfo
  {pages} {93 } (\bibinfo {year} {2000})}\BibitemShut {NoStop}%
\bibitem [{\citenamefont {Baranger}\ and\ \citenamefont
  {Kumar}(1968{\natexlab{a}})}]{BK68-1}%
  \BibitemOpen
  \bibfield  {author} {\bibinfo {author} {\bibfnamefont {M.}~\bibnamefont
  {Baranger}}\ and\ \bibinfo {author} {\bibfnamefont {K.}~\bibnamefont
  {Kumar}},\ }\href {\doibase http://dx.doi.org/10.1016/0375-9474(68)90370-9}
  {\bibfield  {journal} {\bibinfo  {journal} {Nuclear Physics A}\ }\textbf
  {\bibinfo {volume} {110}},\ \bibinfo {pages} {490 } (\bibinfo {year}
  {1968}{\natexlab{a}})}\BibitemShut {NoStop}%
\bibitem [{\citenamefont {Baranger}\ and\ \citenamefont
  {Kumar}(1968{\natexlab{b}})}]{BK68-2}%
  \BibitemOpen
  \bibfield  {author} {\bibinfo {author} {\bibfnamefont {M.}~\bibnamefont
  {Baranger}}\ and\ \bibinfo {author} {\bibfnamefont {K.}~\bibnamefont
  {Kumar}},\ }\href {\doibase http://dx.doi.org/10.1016/0375-9474(68)90044-4}
  {\bibfield  {journal} {\bibinfo  {journal} {Nuclear Physics A}\ }\textbf
  {\bibinfo {volume} {122}},\ \bibinfo {pages} {241 } (\bibinfo {year}
  {1968}{\natexlab{b}})}\BibitemShut {NoStop}%
\bibitem [{\citenamefont {Yuldashbaeva}\ \emph {et~al.}(1999)\citenamefont
  {Yuldashbaeva}, \citenamefont {Libert}, \citenamefont {Quentin},\ and\
  \citenamefont {Girod}}]{YLQG99}%
  \BibitemOpen
  \bibfield  {author} {\bibinfo {author} {\bibfnamefont {E.~K.}\ \bibnamefont
  {Yuldashbaeva}}, \bibinfo {author} {\bibfnamefont {J.}~\bibnamefont
  {Libert}}, \bibinfo {author} {\bibfnamefont {P.}~\bibnamefont {Quentin}}, \
  and\ \bibinfo {author} {\bibfnamefont {M.}~\bibnamefont {Girod}},\ }\href
  {\doibase http://dx.doi.org/10.1016/S0370-2693(99)00836-9} {\bibfield
  {journal} {\bibinfo  {journal} {Physics Letters B}\ }\textbf {\bibinfo
  {volume} {461}},\ \bibinfo {pages} {1 } (\bibinfo {year} {1999})}\BibitemShut
  {NoStop}%
\bibitem [{\citenamefont {Bohr}\ and\ \citenamefont {Mottelson}(1975)}]{BM75}%
  \BibitemOpen
  \bibfield  {author} {\bibinfo {author} {\bibfnamefont {A.}~\bibnamefont
  {Bohr}}\ and\ \bibinfo {author} {\bibfnamefont {B.~R.}\ \bibnamefont
  {Mottelson}},\ }\href@noop {} {\emph {\bibinfo {title} {Nuclear Structure,
  Vol. II}}}\ (\bibinfo  {publisher} {W. A. Benjamin},\ \bibinfo {address} {New
  York},\ \bibinfo {year} {1975})\BibitemShut {NoStop}%
\bibitem [{\citenamefont {Wen}\ and\ \citenamefont
  {Nakatsukasa}(2016)}]{WenN16}%
  \BibitemOpen
  \bibfield  {author} {\bibinfo {author} {\bibfnamefont {K.}~\bibnamefont
  {Wen}}\ and\ \bibinfo {author} {\bibfnamefont {T.}~\bibnamefont
  {Nakatsukasa}},\ }\href {\doibase 10.1103/PhysRevC.94.054618} {\bibfield
  {journal} {\bibinfo  {journal} {Phys. Rev. C}\ }\textbf {\bibinfo {volume}
  {94}},\ \bibinfo {pages} {054618} (\bibinfo {year} {2016})}\BibitemShut
  {NoStop}%
\bibitem [{\citenamefont {Wen}\ and\ \citenamefont
  {Nakatsukasa}(2017)}]{WenN17}%
  \BibitemOpen
  \bibfield  {author} {\bibinfo {author} {\bibfnamefont {K.}~\bibnamefont
  {Wen}}\ and\ \bibinfo {author} {\bibfnamefont {T.}~\bibnamefont
  {Nakatsukasa}},\ }\href {\doibase 10.1103/PhysRevC.96.014610} {\bibfield
  {journal} {\bibinfo  {journal} {Phys. Rev. C}\ }\textbf {\bibinfo {volume}
  {96}},\ \bibinfo {pages} {014610} (\bibinfo {year} {2017})}\BibitemShut
  {NoStop}%
\bibitem [{\citenamefont {Thouless}\ and\ \citenamefont
  {Valatin}(1962)}]{TV62}%
  \BibitemOpen
  \bibfield  {author} {\bibinfo {author} {\bibfnamefont {D.~J.}\ \bibnamefont
  {Thouless}}\ and\ \bibinfo {author} {\bibfnamefont {J.~G.}\ \bibnamefont
  {Valatin}},\ }\href@noop {} {\bibfield  {journal} {\bibinfo  {journal} {Nucl.
  Phys.}\ }\textbf {\bibinfo {volume} {31}},\ \bibinfo {pages} {211} (\bibinfo
  {year} {1962})}\BibitemShut {NoStop}%
\bibitem [{\citenamefont {Hinohara}(2015)}]{Hin15}%
  \BibitemOpen
  \bibfield  {author} {\bibinfo {author} {\bibfnamefont {N.}~\bibnamefont
  {Hinohara}},\ }\href {\doibase 10.1103/PhysRevC.92.034321} {\bibfield
  {journal} {\bibinfo  {journal} {Phys. Rev. C}\ }\textbf {\bibinfo {volume}
  {92}},\ \bibinfo {pages} {034321} (\bibinfo {year} {2015})}\BibitemShut
  {NoStop}%
\bibitem [{\citenamefont {Baran}\ \emph {et~al.}(2011)\citenamefont {Baran},
  \citenamefont {Sheikh}, \citenamefont {Dobaczewski}, \citenamefont
  {Nazarewicz},\ and\ \citenamefont {Staszczak}}]{Bar11}%
  \BibitemOpen
  \bibfield  {author} {\bibinfo {author} {\bibfnamefont {A.}~\bibnamefont
  {Baran}}, \bibinfo {author} {\bibfnamefont {J.~A.}\ \bibnamefont {Sheikh}},
  \bibinfo {author} {\bibfnamefont {J.}~\bibnamefont {Dobaczewski}}, \bibinfo
  {author} {\bibfnamefont {W.}~\bibnamefont {Nazarewicz}}, \ and\ \bibinfo
  {author} {\bibfnamefont {A.}~\bibnamefont {Staszczak}},\ }\href {\doibase
  10.1103/PhysRevC.84.054321} {\bibfield  {journal} {\bibinfo  {journal} {Phys.
  Rev. C}\ }\textbf {\bibinfo {volume} {84}},\ \bibinfo {pages} {054321}
  (\bibinfo {year} {2011})}\BibitemShut {NoStop}%
\bibitem [{\citenamefont {Inglis}(1954)}]{Ing54}%
  \BibitemOpen
  \bibfield  {author} {\bibinfo {author} {\bibfnamefont {D.~R.}\ \bibnamefont
  {Inglis}},\ }\href {\doibase 10.1103/PhysRev.96.1059} {\bibfield  {journal}
  {\bibinfo  {journal} {Phys. Rev.}\ }\textbf {\bibinfo {volume} {96}},\
  \bibinfo {pages} {1059} (\bibinfo {year} {1954})}\BibitemShut {NoStop}%
\bibitem [{\citenamefont {Nakatsukasa}\ \emph {et~al.}(2007)\citenamefont
  {Nakatsukasa}, \citenamefont {Inakura},\ and\ \citenamefont
  {Yabana}}]{NIY07}%
  \BibitemOpen
  \bibfield  {author} {\bibinfo {author} {\bibfnamefont {T.}~\bibnamefont
  {Nakatsukasa}}, \bibinfo {author} {\bibfnamefont {T.}~\bibnamefont
  {Inakura}}, \ and\ \bibinfo {author} {\bibfnamefont {K.}~\bibnamefont
  {Yabana}},\ }\href@noop {} {\bibfield  {journal} {\bibinfo  {journal} {Phys.
  Rev. C}\ }\textbf {\bibinfo {volume} {76}},\ \bibinfo {pages} {024318}
  (\bibinfo {year} {2007})}\BibitemShut {NoStop}%
\bibitem [{\citenamefont {Inakura}\ \emph {et~al.}(2009)\citenamefont
  {Inakura}, \citenamefont {Nakatsukasa},\ and\ \citenamefont
  {Yabana}}]{INY09}%
  \BibitemOpen
  \bibfield  {author} {\bibinfo {author} {\bibfnamefont {T.}~\bibnamefont
  {Inakura}}, \bibinfo {author} {\bibfnamefont {T.}~\bibnamefont
  {Nakatsukasa}}, \ and\ \bibinfo {author} {\bibfnamefont {K.}~\bibnamefont
  {Yabana}},\ }\href {\doibase 10.1103/PhysRevC.80.044301} {\bibfield
  {journal} {\bibinfo  {journal} {Phys. Rev. C}\ }\textbf {\bibinfo {volume}
  {80}},\ \bibinfo {pages} {044301} (\bibinfo {year} {2009})}\BibitemShut
  {NoStop}%
\bibitem [{\citenamefont {Avogadro}\ and\ \citenamefont
  {Nakatsukasa}(2011)}]{AN11}%
  \BibitemOpen
  \bibfield  {author} {\bibinfo {author} {\bibfnamefont {P.}~\bibnamefont
  {Avogadro}}\ and\ \bibinfo {author} {\bibfnamefont {T.}~\bibnamefont
  {Nakatsukasa}},\ }\href {\doibase 10.1103/PhysRevC.84.014314} {\bibfield
  {journal} {\bibinfo  {journal} {Phys. Rev. C}\ }\textbf {\bibinfo {volume}
  {84}},\ \bibinfo {pages} {014314} (\bibinfo {year} {2011})}\BibitemShut
  {NoStop}%
\bibitem [{\citenamefont {Avogadro}\ and\ \citenamefont
  {Nakatsukasa}(2013)}]{AN13}%
  \BibitemOpen
  \bibfield  {author} {\bibinfo {author} {\bibfnamefont {P.}~\bibnamefont
  {Avogadro}}\ and\ \bibinfo {author} {\bibfnamefont {T.}~\bibnamefont
  {Nakatsukasa}},\ }\href {\doibase 10.1103/PhysRevC.87.014331} {\bibfield
  {journal} {\bibinfo  {journal} {Phys. Rev. C}\ }\textbf {\bibinfo {volume}
  {87}},\ \bibinfo {pages} {014331} (\bibinfo {year} {2013})}\BibitemShut
  {NoStop}%
\bibitem [{\citenamefont {Davies}\ \emph {et~al.}(1980)\citenamefont {Davies},
  \citenamefont {Flocard}, \citenamefont {Krieger},\ and\ \citenamefont
  {Weiss}}]{DFKW80}%
  \BibitemOpen
  \bibfield  {author} {\bibinfo {author} {\bibfnamefont {K.~T.~R.}\
  \bibnamefont {Davies}}, \bibinfo {author} {\bibfnamefont {H.}~\bibnamefont
  {Flocard}}, \bibinfo {author} {\bibfnamefont {S.}~\bibnamefont {Krieger}}, \
  and\ \bibinfo {author} {\bibfnamefont {M.~S.}\ \bibnamefont {Weiss}},\
  }\href@noop {} {\bibfield  {journal} {\bibinfo  {journal} {Nucl. Phys. A}\
  }\textbf {\bibinfo {volume} {342}},\ \bibinfo {pages} {111} (\bibinfo {year}
  {1980})}\BibitemShut {NoStop}%
\bibitem [{\citenamefont {Bonche}\ \emph {et~al.}(1976)\citenamefont {Bonche},
  \citenamefont {Koonin},\ and\ \citenamefont {Negele}}]{BKN76}%
  \BibitemOpen
  \bibfield  {author} {\bibinfo {author} {\bibfnamefont {P.}~\bibnamefont
  {Bonche}}, \bibinfo {author} {\bibfnamefont {S.}~\bibnamefont {Koonin}}, \
  and\ \bibinfo {author} {\bibfnamefont {J.~W.}\ \bibnamefont {Negele}},\
  }\href@noop {} {\bibfield  {journal} {\bibinfo  {journal} {Phys. Rev. C}\
  }\textbf {\bibinfo {volume} {13}},\ \bibinfo {pages} {1226} (\bibinfo {year}
  {1976})}\BibitemShut {NoStop}%
\bibitem [{\citenamefont {Wen}\ and\ \citenamefont
  {Nakatsukasa}(2020)}]{WenN20}%
  \BibitemOpen
  \bibfield  {author} {\bibinfo {author} {\bibfnamefont {K.}~\bibnamefont
  {Wen}}\ and\ \bibinfo {author} {\bibfnamefont {T.}~\bibnamefont
  {Nakatsukasa}},\ }\href {\doibase 10.3389/fphy.2020.00016} {\bibfield
  {journal} {\bibinfo  {journal} {Frontiers in Physics}\ }\textbf {\bibinfo
  {volume} {8}},\ \bibinfo {pages} {16} (\bibinfo {year} {2020})}\BibitemShut
  {NoStop}%
\bibitem [{\citenamefont {Reinhard}\ \emph {et~al.}(1984)\citenamefont
  {Reinhard}, \citenamefont {Friedrich}, \citenamefont {Goeke}, \citenamefont
  {Gr\"ummer},\ and\ \citenamefont {Gross}}]{RFGGG84}%
  \BibitemOpen
  \bibfield  {author} {\bibinfo {author} {\bibfnamefont {P.~G.}\ \bibnamefont
  {Reinhard}}, \bibinfo {author} {\bibfnamefont {J.}~\bibnamefont {Friedrich}},
  \bibinfo {author} {\bibfnamefont {K.}~\bibnamefont {Goeke}}, \bibinfo
  {author} {\bibfnamefont {F.}~\bibnamefont {Gr\"ummer}}, \ and\ \bibinfo
  {author} {\bibfnamefont {D.~H.~E.}\ \bibnamefont {Gross}},\ }\href {\doibase
  10.1103/PhysRevC.30.878} {\bibfield  {journal} {\bibinfo  {journal} {Phys.
  Rev. C}\ }\textbf {\bibinfo {volume} {30}},\ \bibinfo {pages} {878} (\bibinfo
  {year} {1984})}\BibitemShut {NoStop}%
\end{thebibliography}%


\end{document}